# Experimental setup for testing nanocalorimeter sensors as a plasma diagnostics tool

Carles Corbella[1,2,a], Feng Yi[1], Andrei Kolmakov[3]

[1] Materials Measurement Science Division, MML, NIST, Gaithersburg, MD 20899, USA

[2] Department of Chemistry & Biochemistry, University of Maryland, College Park, MD 20742, USA

[3] Nanoscale Device Characterization Division, PML, NIST, Gaithersburg, MD 20899, USA

[a] Corresponding author. Email: carles.corbellaroca@nist.gov

ORCiDs: CC: 0000-0002-6201-0680; FY: 0000-0002-8269-3685; AK: 0000-0001-5299-4121

The application of microfabricated nanocalorimeters for real-time, highly sensitive, and selective monitoring of the energy fluxes from low-pressure plasma discharges has been reported recently. Here, we describe the experimental setup for complementary characterization of nanocalorimeters' performance using plasma diagnostics techniques. Langmuir probe, retarding field energy analyzer (RFEA), optical emission spectroscopy (OES), quadrupole mass spectrometry (QMS), and thermocouple-based thermometry constitute the plasma monitor suite. We discuss two application examples of the plasma monitor suite: (1) discrimination between electron- and ion-induced heat fluxes via sensor direct current (DC) biasing, and (2) detection of fast temperature transients in pulsed discharges. The main goal of this work is to demonstrate a nanocalorimetry platform as a real-time metrology for plasma processing and process control.

# I. INTRODUCTION

Cold plasma discharges, also termed technical plasmas, are widely used in the industry for material processing[1-6]. The thermal load on surfaces in contact with non-thermal plasma discharges is an important factor in process control, as it governs the reactions of many physicochemical processes occurring during plasma-wafer interaction. A direct measurement of temperature from plasma-induced thermal loads is usually performed using heat flux probes, which transduce the incident energy flux from the plasma into an electrical signal[7-9].

Calorimetry is an important addition to the traditional plasma diagnostics toolkit, which commonly consists of electrical probes[9,10], optical spectroscopies[11,12], and mass spectrometry[13], since it directly probes the heat balance in plasmas and sample-plasma interfaces. Many different types and designs of calorimeters have been proposed for plasma research, depending on the type of plasma, physical principle of operation, sensor size, and/or type of signal readout. These include pyroelectric calorimeters[14], resistors (thermistors), and thermocouples (thermopiles) attached to a probe surface[15,16]. In general, passive thermal probes detect temperature changes in a calibrated sample exposed to cold plasma. The sensitivity of such probes is limited by the thermal mass of the device or addenda, which amounts from circa 1 J/K, typical for centimeter-sized sensors, down to 0.01 J/K for miniature thermocouples and thin-film thermistors[17-19]. Table I compares the main heat sensor types for real-time, in situ nonthermal plasma calorimetry reported in the literature: from Gardon structures[20] to micro-sized sensors[21]. It is worth noting that technological demands for industrial applications, such as magnetron sputtering deposition[22,23] and plasma etching for microfabrication[24], have led to the development of calorimetric/heat flux probes with sensitivity of a few $W/m^2$ and time resolution down to milliseconds.

TABLE I. Characteristics and typical performance parameters of the main probe types used for cold plasma calorimetry measurements, along with relevant references from the literature.

| **Probe type** | **Transducer type** | **Calibration** | **Typical sensor size ($mm^2$)** | **Energy flux range ($W/m^2$)** | **Resp. time (s)** | **Refs.** |
|---|---|---|---|---|---|---|
| Calorimetric probe | Thermocouple | Temperature | 300 | $< 10$ to $< 10^4$ | $\approx 1$ | 8,17,18 |
| Gardon heat flux sensor | Thermocouple | Energy flux | 100 | $\approx 10$ to $< 10^4$ | $< 1$ | 20,24 |
| Heat flux microsensor | Thermopile | Energy flux | < 100 | $\approx 1$ to $< 10^4$ | $< 10^{-3}$ | 23,25,26 |
| Nanocal. Sensor | Thin-film thermistor | Temperature | < 10 | $\approx 10$ to $\approx 10^3$ | $< 10^{-3}$ | 21,27 |

Microfabricated nanocalorimeters feature small thermal mass, simple design, and use standard micro/nano-electromechanical systems (MEMS/NEMS) lithography techniques for fabrication[28-30]. Consequently, they can sensitively measure rapid heating and cooling cycles to quantify the incoming heat fluxes or heat generated during fast surface reactions and transformations. This capability offers key advantages for plasma diagnostics, including high sensitivity, excellent selectivity, and exceptionally fast response times of less than 0.1 ms. The initial motivation for designing nanocalorimeters in 1990s was the development of the fast scanning (differential) calorimetry for ultrathin materials and their thermoanalytical modeling[31-33]. More recently, such sensors have been used to measure surface chemical processes and phase transitions, including those involving nanomaterials[27,34]. Due to their excellent mechanical and thermal properties, ultrathin $SiN_x$ membranes supported by Si frames are often used in nanocalorimeters[35-37]. These design features offer a number of advantages. First, fast responses are due to the reduced heat capacity of the sensing element and its thermal isolation from the supporting Si frame. Second, the nanocalorimeter's sensitivity, defined as the ratio of heat flux to the corresponding measured temperature increase, can be as low as 1 nJ/K. Such a value is much smaller than that of traditional probes with larger thermal masses[38,39]. Third, the chemical selectivity to the plasma reactive species or charged particles can be achieved by depositing catalyst coatings and/or adding extra functionalities, e.g., a reference sensor as in differential nanocalorimetry[40], or applying direct current (DC) bias[21], pre-setting temperature or temperature ramping, or a combination thereof. Using these selectivity approaches, the heat flux partitioning from different plasma species, such as ions, electrons, neutrals and photons, can be achieved.

The success and reach of nanocalorimeter metrology in the characterization of chemical processes and material properties at the nanoscale can be gauged by a sustained innovation activity[41-43], which has enabled their integration into other metrologies. For instance, the sensor design can be adapted for thermal monitoring in electron microscopy and other microanalysis techniques[44,45]. Additionally, nanocalorimeter sensors can be combined with standard plasma probes, yielding versatile plasma-monitoring systems. For instance, efforts to integrate calorimetry with the Langmuir probe[19] and retarding field energy analyzer (RFEA)[46] have been reported as promising methods for evaluating plasma energy fluxes alongside discharge parameters. Also importantly, differential nanocalorimetry has been used to rapidly measure H radical concentration from hydrogen discharges, thereby demonstrating its sensitivity to neutral particles, in contrast to electrostatic probes[40,47,48]. Recently, the energy fluxes of plasma ions and electrons were measured discriminatively using a DC-biased nanocalorimeter[21]. The advancement of MEMS-based nanocalorimetry for plasma characterization motivates this article.

In this report, we describe the experimental setup and plasma monitor suite designed to test a microfabricated nanocalorimeter as a prospective plasma diagnostic tool. The system was briefly introduced in a previous article on the application of nanocalorimetry to plasma energy flux metrology[21], but its design and capabilities have not yet been described in detail. This experimental setup is designed to establish a direct link between nanocalorimetry data and the thermal, electrical, chemical, and optical properties of discharges, which is crucial for interpreting the underlying thermal processes in plasmas. The suite includes a Langmuir probe, RFEA, optical emission spectroscopy (OES) probe, quadrupole mass spectrometer (QMS), and a thermocouple probe. Their use to characterize plasma discharges along with plasma nanocalorimetry is illustrated in two application examples: separation of ion and electron contributions to the energy flux from inductive discharges, and characterization of thermal transients observed during pulsed plasma operation.

## II. INSTRUMENT SETUP

This section summarizes the main features of the vacuum chamber, the plasma diagnostics suite, and the nanocalorimeter sensor system. The performance of the latter is supported by measurements of plasma parameters, such as plasma potential, ion density and flux, and gas temperature profile. A methodology to interpret the nanocalorimeter's measurements is also included.

### *A. Plasma reactor*

The core setup (Figure 1) consists of a commercial 6-way cross chamber vacuum system with DN 100 CF flanges. One of the horizontal flanges is occupied by a magnetic manipulator that drives an L-shaped sensor holder that positions and rotates the sensor ±90 degrees along a horizontal axis. The holder is made of aluminum and is connected to the reactor mass via a Cu-braided strap for both electrical grounding and thermal sink purposes. Sensor mounting and replacement can be performed through the frontal load lock door with a tempered Pyrex viewport, which also serves as a window for optical diagnostics of the discharge. The reactor is pumped by a turbomolecular pump backed by two dry mechanical pumps connected in series. The typical base pressure is below $10^{-3}$ Pa. Ultra-high-purity argon, oxygen, and helium gases (0.99999 mol/mol) have been used in the present study. Gas pressure was measured using a capacitance manometer and a full-range pressure gauge, and it was controlled by varying the inlet gas flow rate and the pumping speed. The nanocalorimetry tests reported hereafter were performed within 5 Pa to 60 Pa gas pressure range, and between 10 W and 80 W of radiofrequency (RF) power. The reported plasma parameters are typical for low-temperature inductively coupled plasma (ICP) and capacitively coupled plasma (CCP) discharges used for cleaning and etching (low ion energy) applications[1] (see also Sec. III).

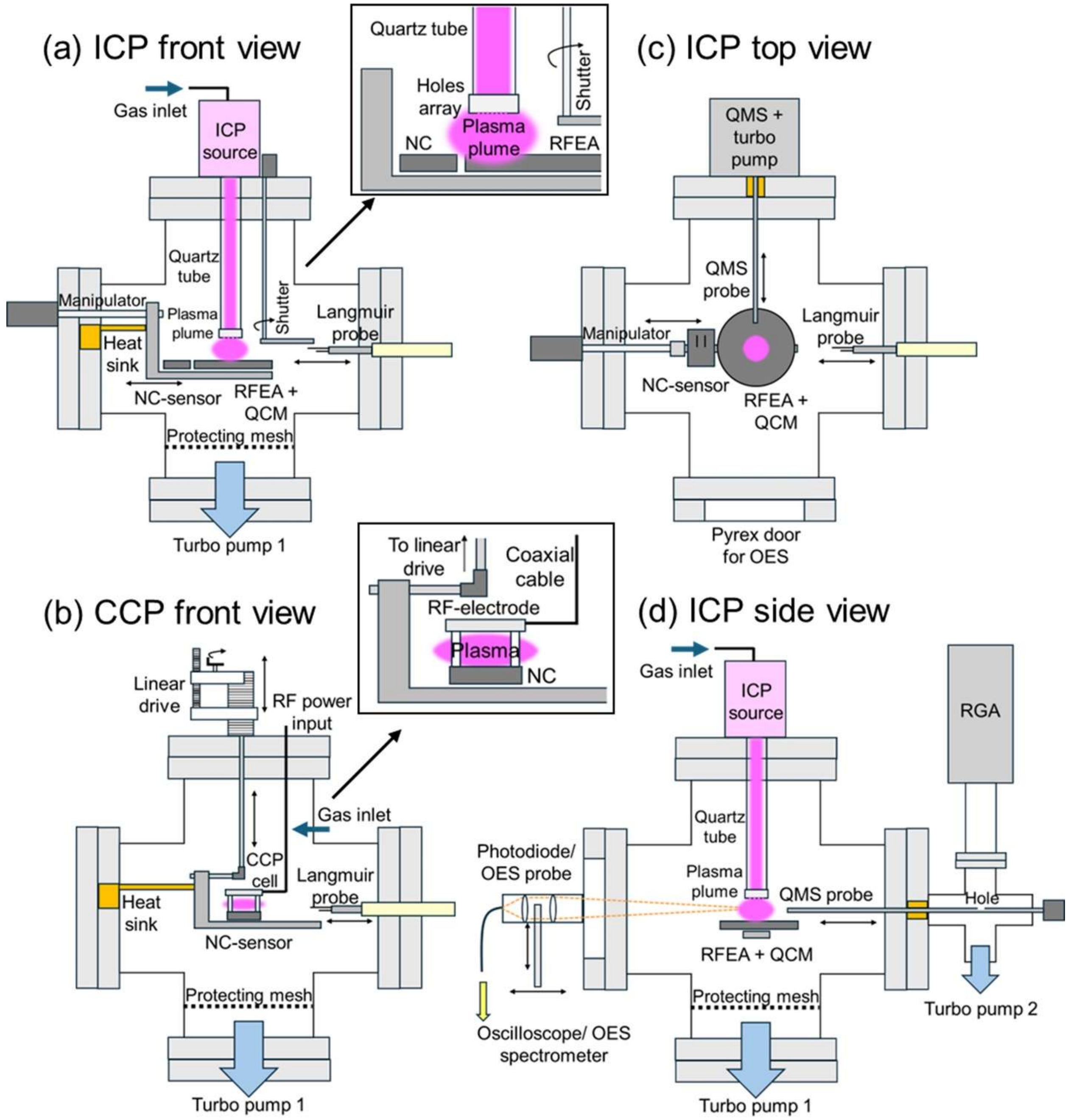

FIG. 1. Cross-sectional schematic views of the plasma reactor for (a, c, d) inductively coupled plasma (ICP) configuration and (b) capacitively coupled plasma (CCP) configuration. The main components are indicated. The setup is equipped with basic plasma diagnostics instruments. NC: Nanocalorimeter. RFEA: Retarding Field Energy Analyzer. QCM: Quartz Crystal Microbalance. OES: Optical Emission Spectroscopy. QMS: Quadrupole Mass Spectrometry. RGA: Residual Gas Analyzer.

The entire assembly, including the pumping units, is easily transportable thanks to its integration into a wheeled frame. In this paper, we report on the capabilities of both plasma configurations, ICP and CCP (Figure 2):

- *ICP configuration:* A remote RF ICP discharge source (13.56 MHz) operating from 10 W to 100 W is installed on the top flange. The generated plasma plume is delivered to the sample holder position through a vertical quartz tube approximately 20 cm long and with 22 mm inner diameter. Its lower end is capped by an electrically grounded Al electrode foil with a set of millimeter-sized holes arranged to improve plasma uniformity at the nozzle. The gap between the nozzle and the sensor surface is 12 mm, and it can be filled with a plasma plume at RF powers of around 50 W or higher. The sensor itself is designed to be electrically floating, allowing it to be biased to a required potential. The plasma flux can be interrupted using a manual shutter.
- *CCP configuration:* In this setup, the sensor is embedded in one of the electrodes of a CCP discharge cell consisting of two parallel plates. The upper plate is an aluminum disc with a diameter of 50 mm and a thickness of 5 mm, which is connected to an external RF power supply (13.56 MHz). The lower plate is supported by the L-shaped sensor holder mentioned above, and it is always grounded. The electrode gap can be varied, but it is fixed to approximately 13 mm in the experiments reported here. It is generally safe operating at power densities below 10 W/cm$^2$ to avoid damaging the CCP source[49], thus the supplied RF power in our experiments was thereby limited to 100 W.

(a) Ar ICP: 6.5 Pa, 80 W

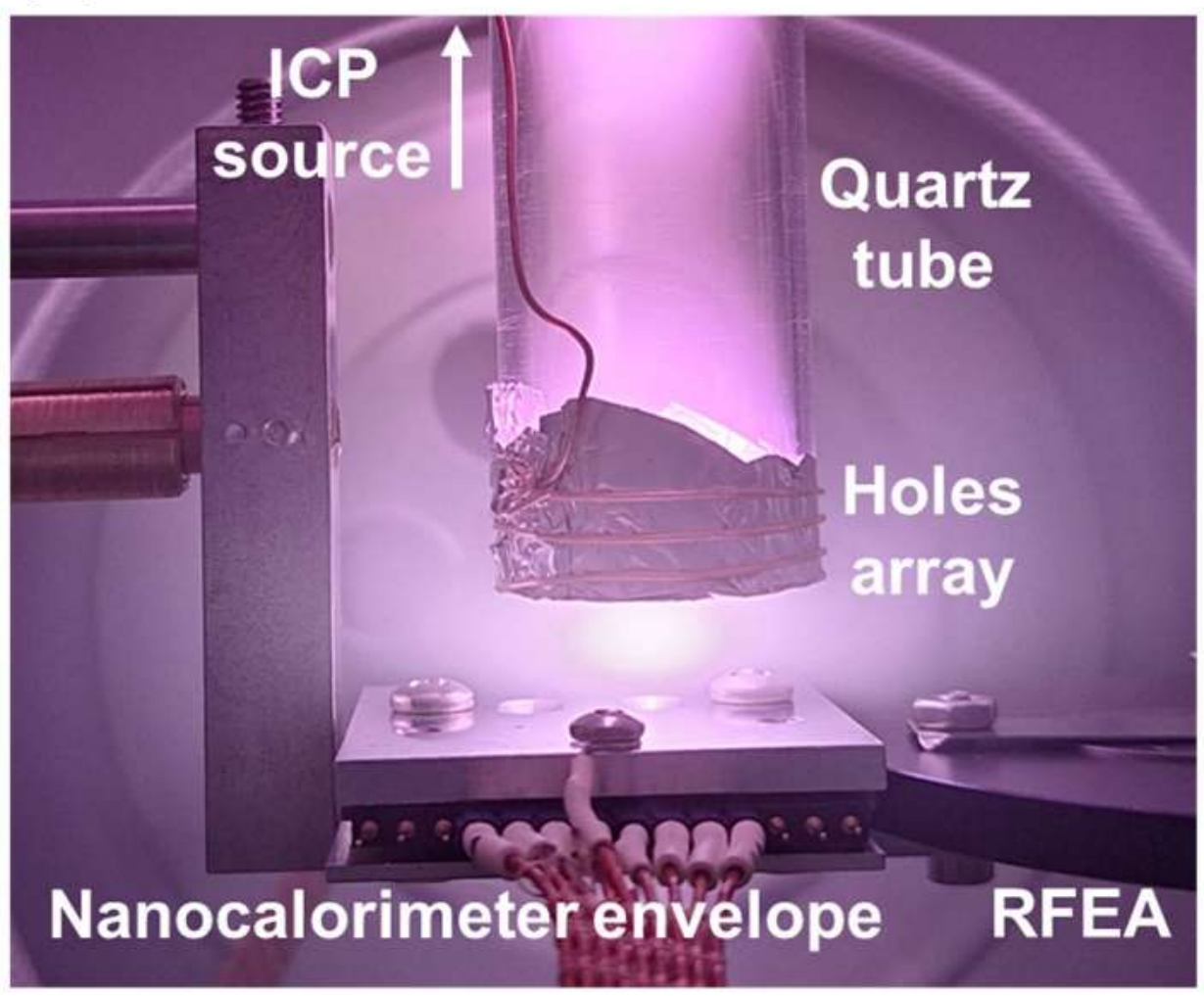

(b) Ar CCP: 13.5 Pa, 60 W

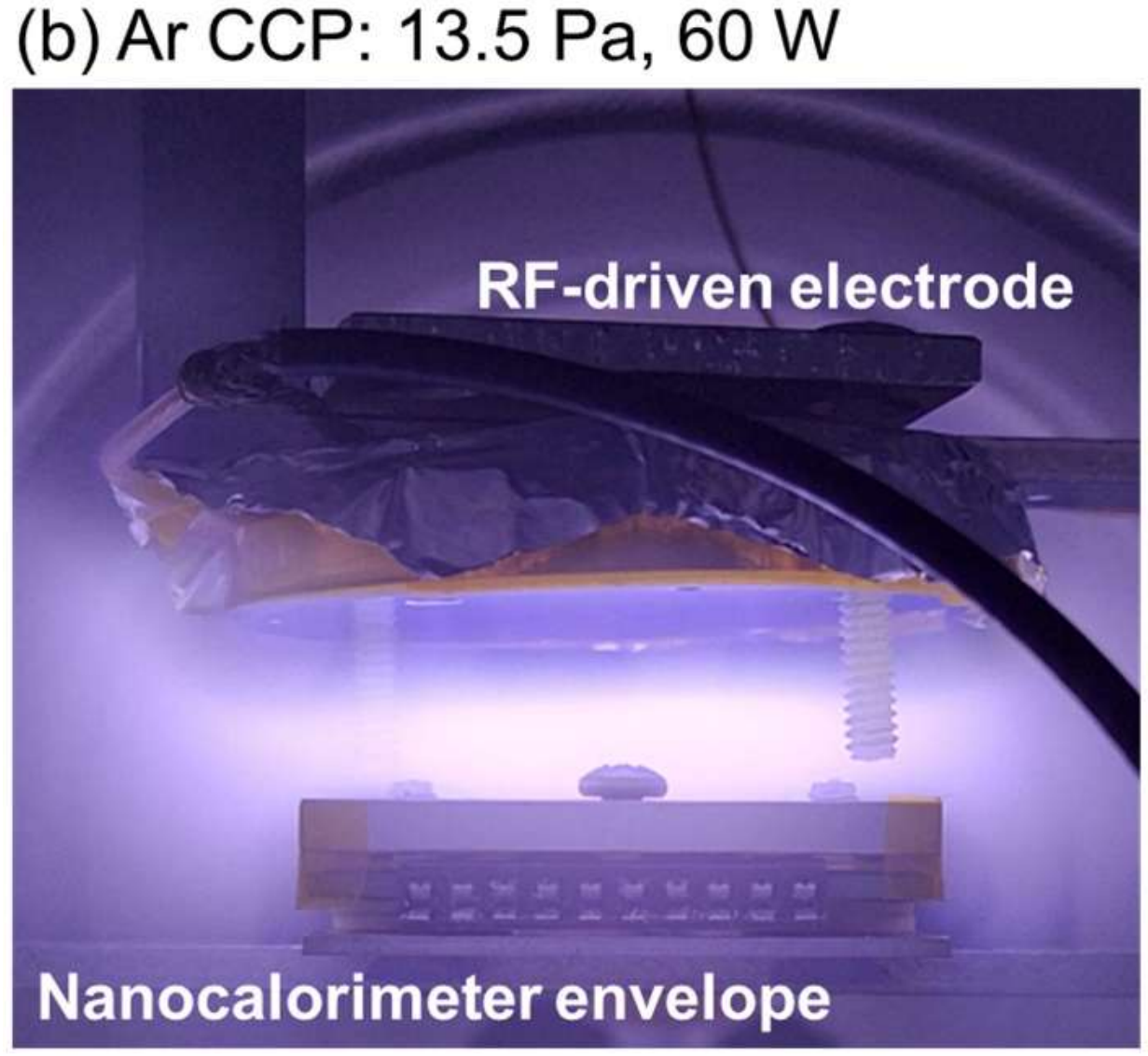

FIG. 2. Images of Ar plasma discharges operated in (a) ICP and (b) CCP configurations. The corresponding discharge parameters are included. ICP: The quartz tube has an external diameter of 25 mm. CCP: The RF-driven electrode has a diameter of 50 mm. The electrical leads to the nanocalorimeter are shielded from electromagnetic fields.

## B. Nanocalorimeter sensor

### 1. Device description

The nanocalorimeter sensor used in our plasma experiments employs a patterned 100 nm thin Pt film thermistor fabricated on a free-standing 100 nm thin $SiN_x$ membrane. Figure 3 shows a schematic cross-section of the sensor structure. Briefly, a lithographically defined Pt strip (100 nm thick) was deposited onto a low-stress $SiN_x$ membrane (100 nm). A Ta interlayer (10 nm) is used to enhance Pt adhesion to $SiN_x$ substrate. Optionally, catalytic coating can be deposited onto the $SiN_x$ side of the sensor to enhance the selective detection of plasma radical species. Further details about chip fabrication are reported elsewhere[50].

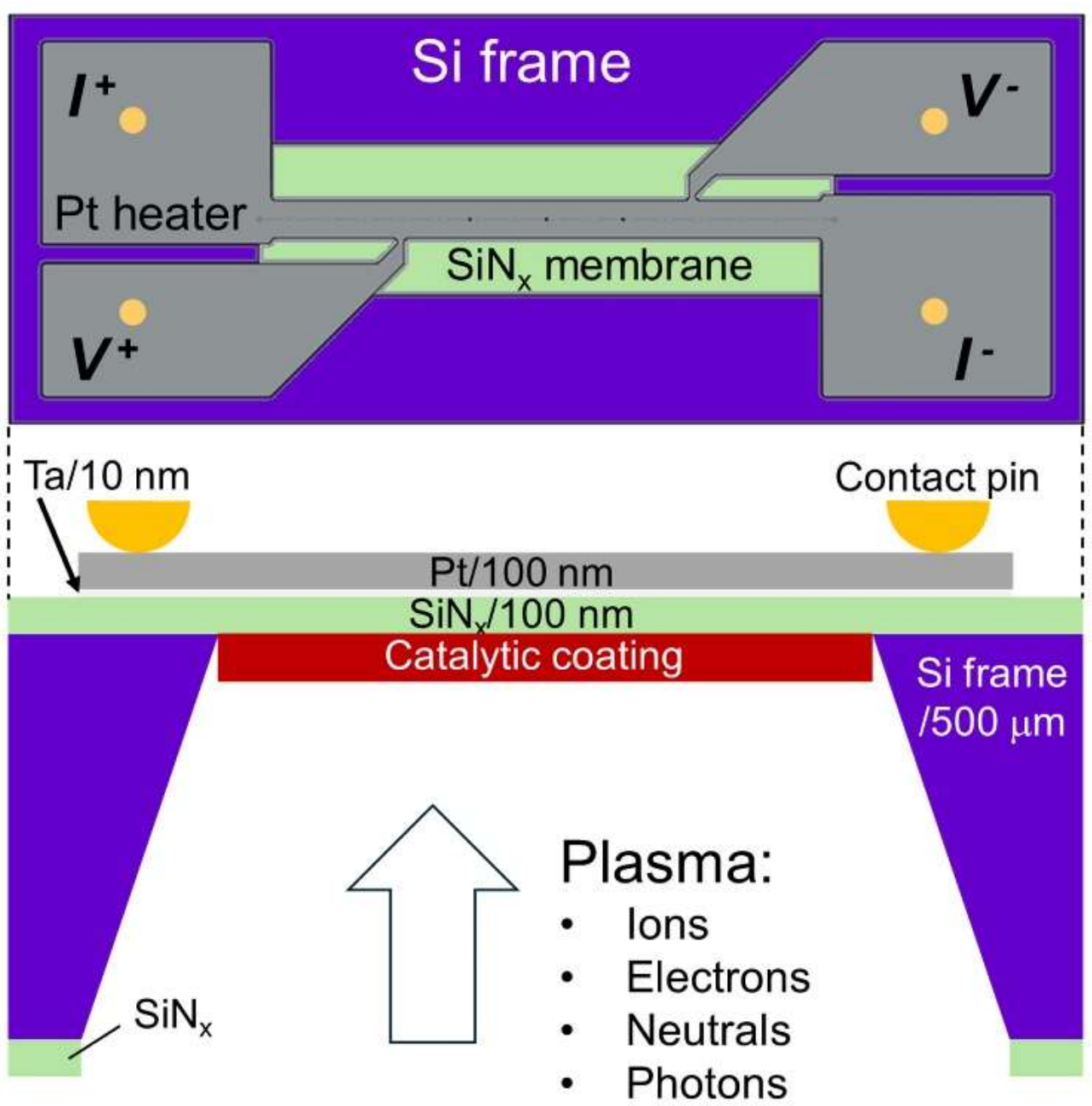

FIG. 3. The nanocalorimeter's top-view and cross-section (not to scale) sketches. The sensor consists of a lithography-defined thin film multilayer stack suspended on a silicon die frame: $SiN_x$ (100 nm)/Ta (10 nm)/Pt (100 nm). Sensitivity to specific elements can be enhanced by depositing a catalytic coating on the plasma-facing side.

After fabrication, the sensors are annealed in air to release the mechanical strain and stabilize the film microstructure. Next, the Pt thermistor is calibrated over the range 293 K to 900 K. The measured temperature coefficient of resistance (TCR) for the sensors was approximately $(3\pm0.2)\times10^{-2}$ Ω/K. The sensor's heat capacity is evaluated using a self-heating-based relaxation method prior to the plasma exposure experiments[21,51]. Briefly, upon a step-up or step-down variation in input current, the time derivative of the sensor temperature and the associated change in power are measured in vacuum. Their ratio provides an effective heat capacity [see Eq. (1)]. An average value of $1.5\times10^{-6}$ J/K for the explored temperature range is obtained, consistent with estimates based on individual material parameters in the stack and prior measurements[52]. See *Supplementary Material* for plots showing resistance-temperature calibration and heat capacity values measured at different temperatures.

## *2. Temperature and energy flux measurements*

The basic electrical setup for the Pt strip temperature measurements in plasma is sketched in Figure 4. The sensor is inserted in an electrically grounded envelope consisting of Al front and rear support plates. The front plate has two rectangular slit openings (0.7 mm × 4.6 mm each, 1 cm apart) for differential nanocalorimetry of plasma using two nanocalorimeters located in adjacent slots. Four spring-loaded (Pogo) pins on a printed circuit board (PCB) contact the Pt heater pads and mechanically press the sensor chip against the front plate. The frame around the openings was rimmed with insulating 0.1 mm thick Kapton film to ensure electrical isolation between the sensor and the grounded enclosure.

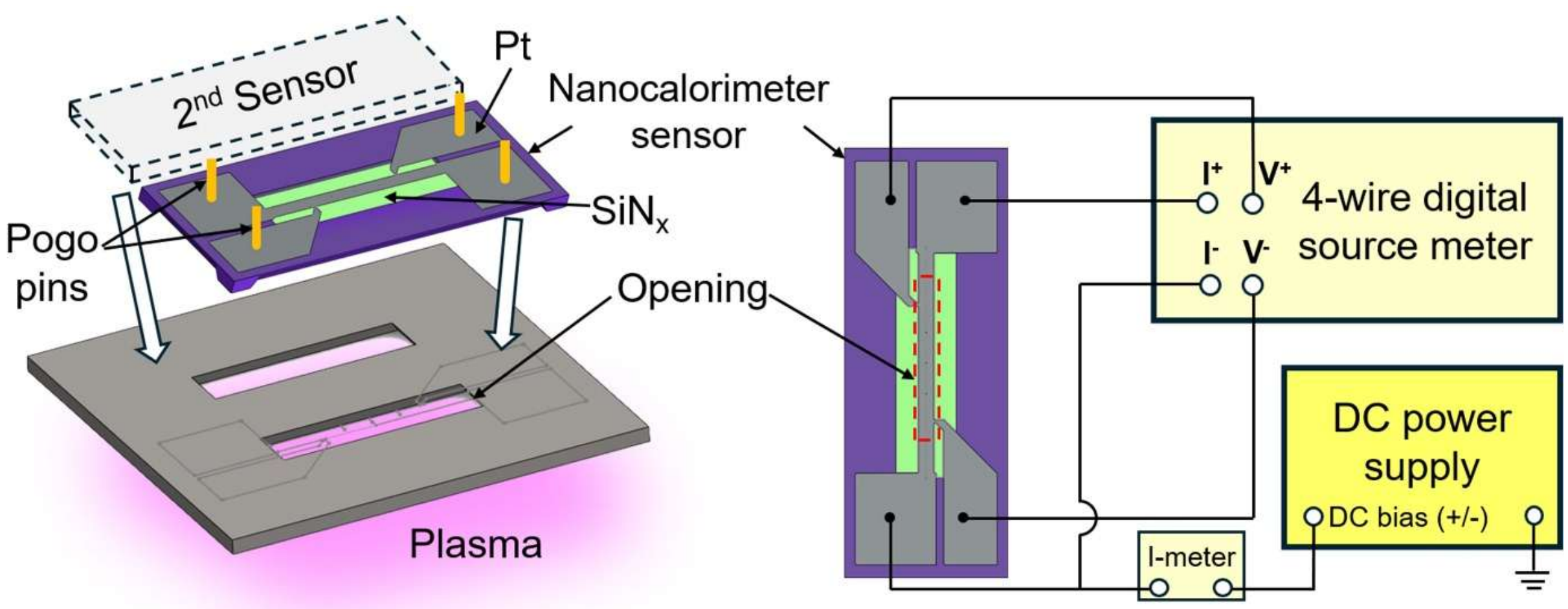

FIG. 4. Nanocalorimeter sensors are exposed to the plasma environment through rectangular slit openings (red dashed rectangle shows the opening area) on the Al enclosure. Only one sensor was used in this article. The second opening can be used for an additional nanocalorimeter for differential calorimetry, or for an alternative temperature sensor, such as a thermocouple or a commercial thermistor. A printed circuit board (not shown here) connects to the nanocalorimeter via spring-loaded (Pogo) pins for resistance measurements. Connections of the source meter unit and DC power supply for 4-probe resistance measurements, plasma current measurements, and additional biasing are also shown.

The PCB is connected to an external source-meter unit, which measures resistance in a four-probe configuration by applying a current and sensing the potential difference across the Pt microstrip. A source current of 1 mA was used to minimize self-heating of the nanocalorimeter. The sensor can be biased to a DC potential with respect to ground by connecting the source unit's low terminal pin (floating) to an external voltage supply.

As indicated above, surfaces exposed to plasma discharges receive the heat flux which consists of contributions from ions, electrons, photons, and neutral species, including reactive radicals and energetic neutrals like metastable atoms. Note that sensitivity to selected plasma agents can be enhanced by depositing a catalytic film or applying an electrical bias. In general, variation of sensor temperature, $T$, can be modeled through a heat balance equation for the nanocalorimeter exposed to a plasma environment:

$$C_\mathrm{s}\frac{\mathrm{d}T}{\mathrm{d}t} = P_\mathrm{in} - P_\mathrm{out} \tag{1}$$

where $C_\mathrm{s}$ is the sensor heat capacity, $P_\mathrm{in}$ is the incident power due to the interaction with the above plasma species, and $P_\mathrm{out}$ is the outgoing power due to thermal losses. An effective $C_\mathrm{s}$ of the Pt thermistor was determined to be approximately $1.5\times10^{-6}$ J/K, as mentioned above. Heat losses caused by thermal radiation and convection can generally be neglected at the temperatures and pressures used in this study. Thus, assuming heat dissipation exclusively by thermal conduction to the Si frame and to the gas, $P_\mathrm{out} = h \cdot \left(T - T_\mathrm{eq}\right)$, where $h$ is a heat transfer coefficient and $T_\mathrm{eq}$ is the temperature of the environment (Al envelope).

Under steady state conditions, i.e., when sensor temperature is constant, $P_\mathrm{in} = P_\mathrm{out}$ from Eq. (1). Any variation of the heat flux to the sensor, $\Delta P$, creates a thermal imbalance that changes temperature. Such energy flux can be estimated by evaluating the time derivative of temperature just after the thermal change is applied, $C_\mathrm{s}\frac{\mathrm{d}T}{\mathrm{d}t} = \Delta P$. This methodology will be applied in Sec. IV to measure energy fluxes from plasma sources.

# III. PLASMA DIAGNOSTICS SUITE

Plasma parameters, temperature, and chemical composition, relevant to nanocalorimetry experiments, have been collected using the plasma characterization suite. This section is structured according to the diagnostic technique employed to characterize the RF discharge and support the nanocalorimetry study. Descriptions of the main hardware for each tool are provided, followed by a short analysis of measurements.

## A. Electrostatic probes: plasma parameter range

### 1. Langmuir probe

Plasma parameters, in particular plasma density, electron temperature, and plasma potential, as well as the electron energy distribution function (EEDF), were measured using a Langmuir probe along with a compensation electrode to avoid I-V curve distortions near plasma potential. Inherent distortions caused by RF potential oscillations were avoided due to inductance filters integrated into the probe's electronic unit. The probe tip consists of a 10 mm long, 0.4 mm diameter tantalum wire, and it can be moved along the same axis as the magnetically driven holder. A double Langmuir probe is also available to alternatively evaluate plasma parameters, independent of the plasma potential. The dependencies of parameters with RF power and pressure reported below help characterize the discharges studied for plasma nanocalorimetry in this setup.

Langmuir probe measurements were performed in Ar ICP discharges at 2.5 Pa and 6.5 Pa, and RF power ranging from 20 W to 90 W. Figure 5 shows an exemplary I-V curve, where main regions of interest are identified. The recorded curve has a characteristic shape, including ion and electron saturation regions at high negative and positive probe voltages, respectively. The electron temperature is inferred from the central retardation region, which includes contributions from both ion and electron currents, assuming a Maxwellian electron distribution. I-V curve analysis using a collisional sheath model was used to calculate the plasma density[53]. Plasma potential is evaluated using the intersecting-slopes method from the retardation and electron-saturation regions.

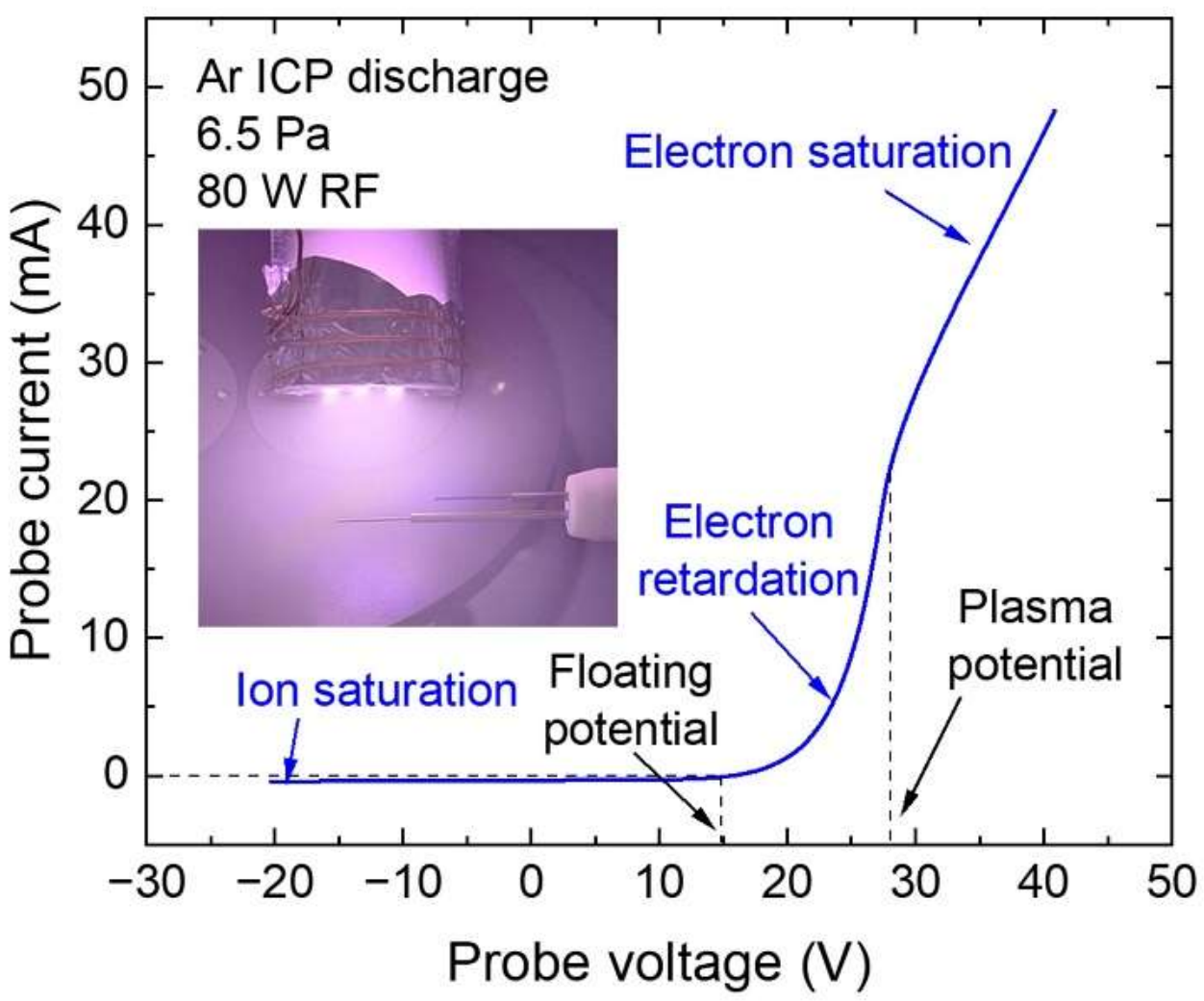

FIG. 5. Typical I-V curve recorded for an Ar ICP discharge using a Langmuir probe. The indicated regions are used to model a Maxwellian electron distribution, from which basic plasma parameters were retrieved (total sampling time per voltage step: 100 µs; trace average per scan: 10; number of scans: 10). Inset: picture of the probe (together with DC compensation probe) exposed to the Ar plasma.

Figure 6 shows the evolution of the plasma potential, electron temperature, and electron/ion or plasma density. These parameters are measured with an instrumental uncertainty of around 1 %. The plasma potential increases slightly from 25 V to 30 V with increasing RF power. The increase in plasma density is more pronounced at 6.5 Pa than at 2.5 Pa, showing maximal values of $4.5\times10^{16}$ $m^{-3}$ and $2.5\times10^{16}$ $m^{-3}$, respectively. Therefore, an increase in pressure leads to a clear increase in plasma density, while the plasma potential is not significantly affected by pressure changes within this range. The electron temperature is practically constant, independent of RF power (3.0 eV to 3.5 eV), and, as expected, it slightly decreases with gas pressure due to collision-induced decreases in the mean electron energy (shorter mean free path).

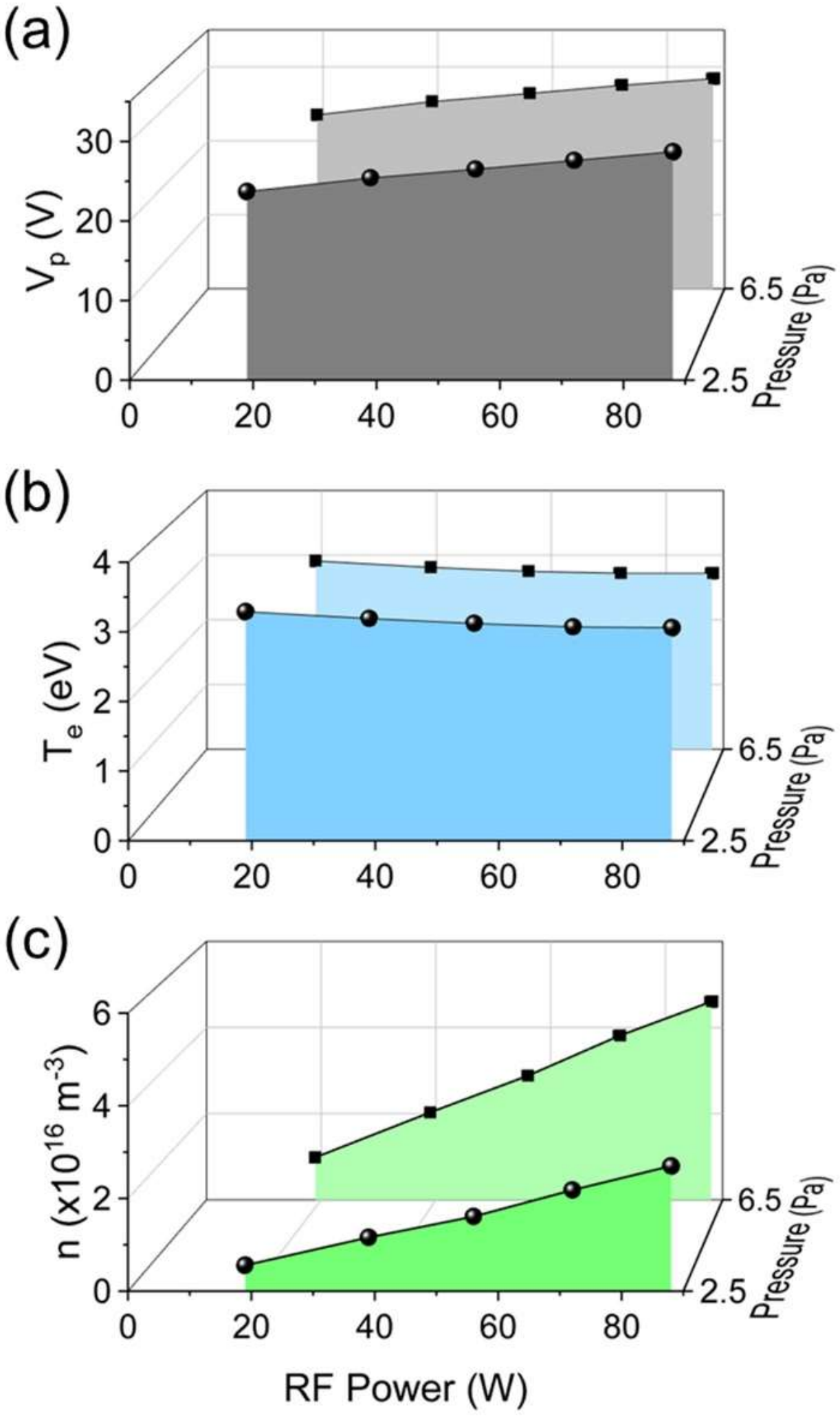

FIG. 6. Dependencies of (a) plasma potential, $V_{\mathrm{p}}$, (b) electron temperature, $T_{\mathrm{e}}$, and (c) plasma density, $n$, measured via Langmuir probe, with varying RF power and gas pressure in an Ar ICP discharge. The instrumental uncertainty is approximately 1 % for each parameter.

Plasma potentials and electron temperatures measured in ICP and CCP systems are of the same order within the same pressure and RF power ranges. However, plasma density values of ICP discharges are roughly twice the CCP values. This shift toward higher densities is natural in inductive discharges compared to their capacitive counterparts[1].

## 2. *Retarding field energy analyzer*

RFEA is a sheath diagnostics instrument that provides the energy distribution and flux of incoming ions. The holder in the vacuum chamber contains a nanocalorimeter enclosure and a 10 cm diameter RFEA housing. A stack of parallel grids prevents plasma electrons from entering the analyzer and discriminates the incoming ions by their kinetic energy as the retarding potential increases. As a result, the ion current vs. retarding voltage is measured and converted into the ion energy distribution function (IEDF). The total ion flux can also be obtained by accounting for the grid system's transmission factor. The collector electrode consists of a built-in quartz crystal microbalance (QCM) to monitor mass variation rates. The distance between the upper grid and the collector is approximately 0.5 mm, minimizing the effect of ion scattering from internal collisions at the working pressures. Ion energies of up to 2000 eV can be collected by externally biasing the RFEA housing. This setup allows IEDF measurements as a function of RF power and pressure, directly correlating ion energies with the plasma potential, which is essential for linking these parameters to recorded heat fluxes.

Figure 7 shows typical IEDF curves measured via RFEA of Ar ICP discharges. The RFEA housing was grounded. Curves with bimodal profiles are observed (Figure 7a) at different pressures. The origin of the two relative maxima is probably related to oscillations of the plasma potential. Consistently, the total ion flux (area under each curve) increases with pressure, corroborating plasma density evolution. The low-energy tail in IEDF is less intense at lower pressures due to less frequent ion scattering within the sheath. At a fixed pressure of 6.5 Pa (Figure 7b), the low-energy tail loses weight when power increases because the plasma sheath becomes thinner, so larger fractions of the total ion flux reach the analyzer without collisions.

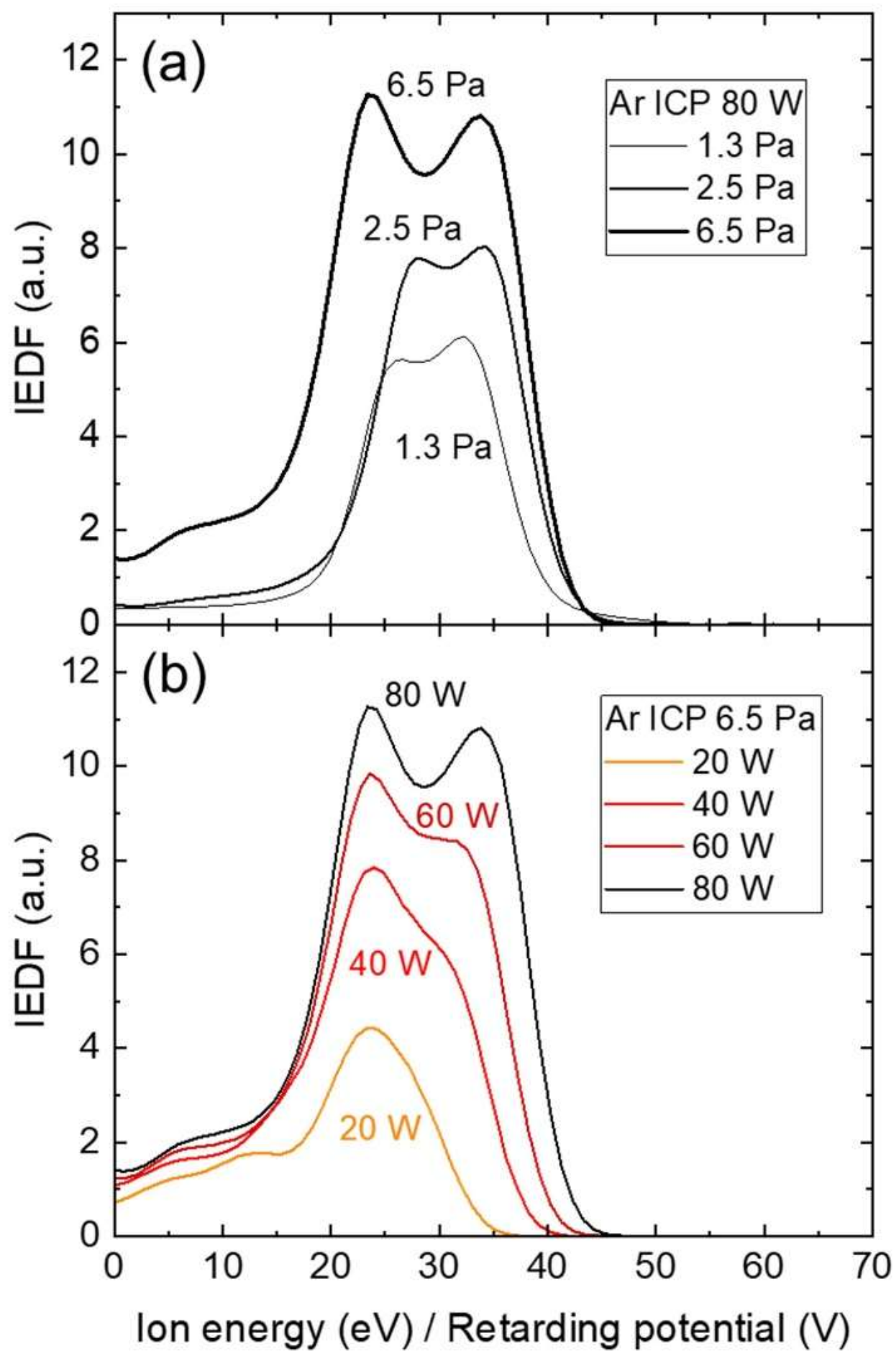

FIG. 7. Ion energy distribution functions (IEDF), measured via retarding field energy analyzer (RFEA), of Ar ICP discharge held (a) at 80 W with varying pressure and (b) at 6.5 Pa with varying RF power (integration time: 10 ms; number of scans: 10).

Figure 8a shows a linear correlation between the average peak ion energy, measured via RFEA, and plasma potential, measured with the Langmuir probe for plasma discharges at different pressures and power values. The observed correlation is not surprising because, for sheaths with low collision rates, most incident ions (singly ionized) accelerate across the sheath, gaining kinetic energy equal to the plasma potential. The collisional regime of the plasma sheath is briefly discussed in the *Supplementary Material*. The ion flux is proportional to RF power, as shown in Figure 8b and described above. This behavior is directly related to the increase in ion density with RF power. In conclusion, the ion current density increases with increasing pressure and/or RF power. The ion flux and ion energy distribution data are crucial to properly characterize nanocalorimetry measurements, as elaborated in Sec. IV.

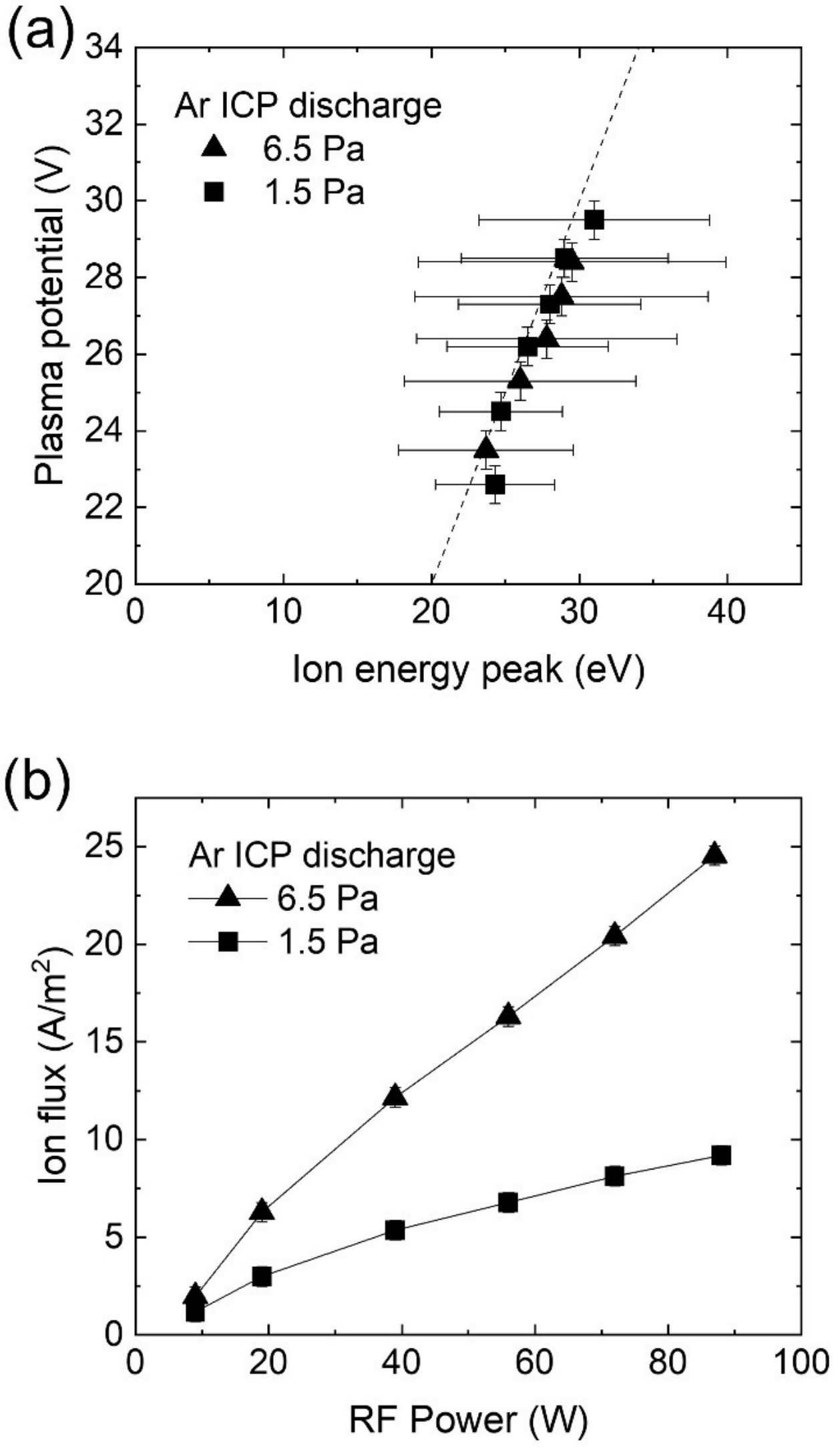

FIG. 8. (a) Correlation between plasma potential (Langmuir probe) and average peak energy of ion kinetic energy (RFEA) for a grounded electrode in Ar ICP discharge. The horizontal error

bars account for the full width at half maximum (FWHM) of each ion energy distribution function (IEDF). The vertical error bars, corresponding to the instrument uncertainty of the Langmuir probe, are limited to ±0.5 V. The dashed straight line is a guide to the eye that intersects the origin of the coordinates. (b) Increase of ion flux (integral of IEDF over retarding potential) with pressure and RF power. The vertical error bars, due to statistical RFEA measurement uncertainty, are on the order of ±0.5 $A/m^2$.

## B. *Optical and mass spectrometries: plasma chemistry*

### 1. *Optical emission spectroscopy*

OES is one of the major techniques for plasma diagnostics, including chemical composition analysis. As mentioned above, light emitted by the plasma discharge was collected through the tempered Pyrex viewport. Two condenser lenses were used to collect light from the center of the discharge into a fiber-optic light guide connected to a UV-Vis optical spectrometer with a spectral range of 188 nm to 1044 nm and 1.3 nm resolution. Alternatively, a silicon photodiode can be connected to the condenser lenses in order to measure total intensity of the optical emission from plasma and for high temporal resolution.

The typical OES spectra of the Ar and $O_2$ generated plasma plume are depicted in Figure 9 together with the literature values of peaks' wavelengths. The main emission lines and corresponding transitions are listed in Table II. The brightest line of argon discharge is located at 811.53 nm (Figure 9c), while the dominant contribution from O atoms in $O_2$ plasma is found at 777.4 nm (Figure 9d). A small peak at around 310 nm within a broad background could be attributed to residual water contamination.

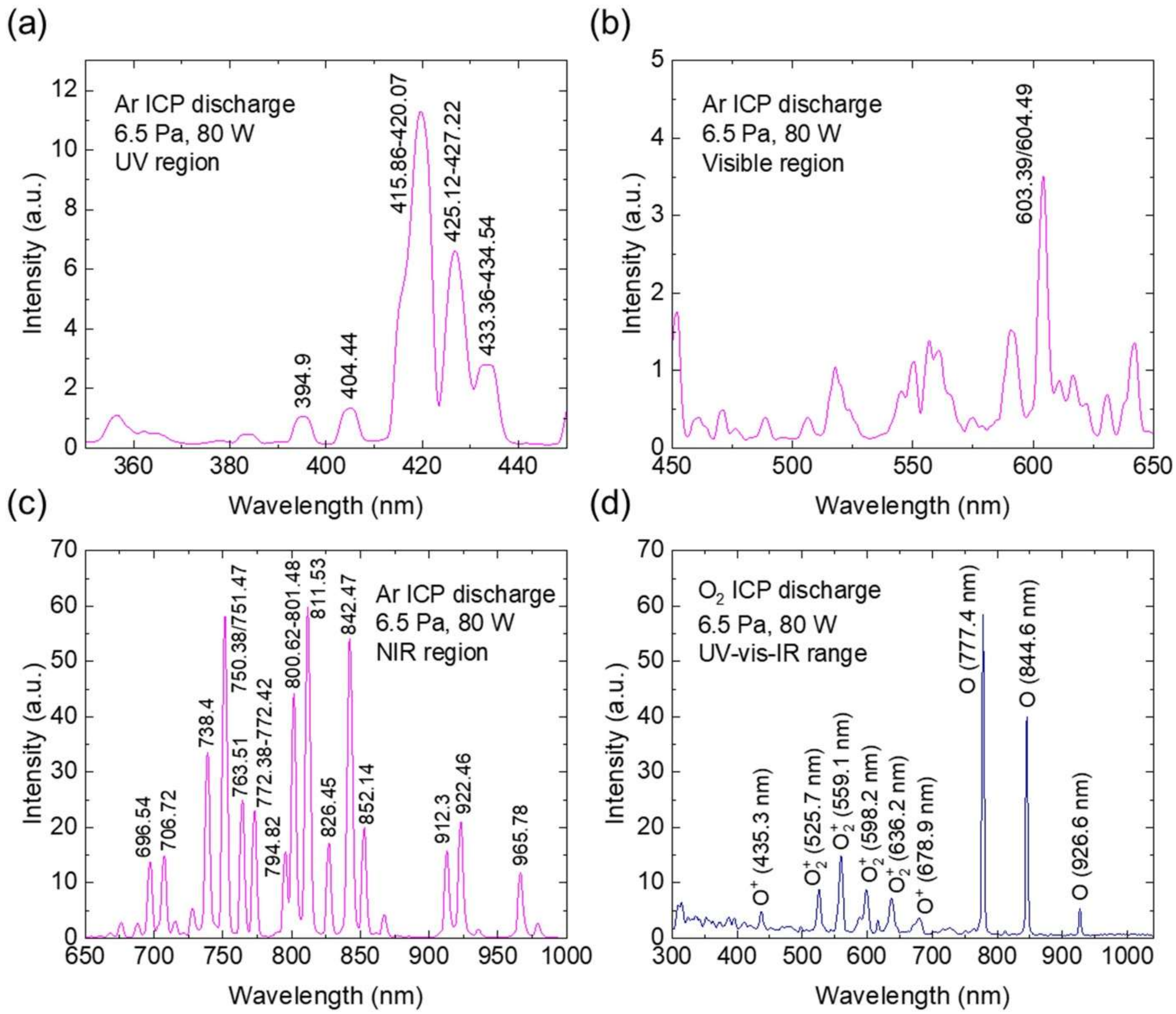

FIG. 9. Optical emission spectroscopy (OES) spectra of Ar ICP discharge in (a) ultraviolet (UV) region, (b) visible region, and (c) near infrared (NIR) region. (d) OES of $O_2$ ICP discharge. Wavelengths (in nm) of the most relevant optical transitions are identified.

TABLE II. Optical transitions of the most intense emission lines observed in the OES spectra of Ar and $O_2$ ICP discharges[54-56]:

| Wavelength | Transition line |
|---|---|
| 750.39 nm Ar I | 4p’ $[0^1/_2]_0 \rightarrow$ 4s’ $[0^1/_2]_1$ |
| 751.47 nm Ar I | 4p $[0^1/_2]_0 \rightarrow$ 4s $[1^1/_2]_1$ |
| 811.53 nm Ar I | 4p $[2^1/_2]_3 \rightarrow$ 4s $[1^1/_2]_2$ |
| 842.46 nm Ar I | 4p $[2^1/_2]_2 \rightarrow$ 4s $[1^1/_2]_1$ |
| 559.1 nm $O_2^+$ | $b_4\Sigma_g^- \rightarrow a_4\Pi_u$ |
| 777.42 nm O I | 3p $^5P \rightarrow$ 3s $^5S^0$ |
| 844.64 nm O I | 3p $^3P \rightarrow$ 3s $^3S^0$ |

Evolutions in emission line intensities in an Ar ICP discharge at 6.5 Pa, measured via OES, while varying the source power from 10 W to 80 W, suggest the existence of two power regimes (Figure 10). The low-power regime extends from low source power up to approximately 40 W, after which the discharge enters a high-power regime characterized by a slower increase in OES line intensity. In the latter region, the smoother increase in line emission with RF power indicates a more efficient excitation/ionization of plasma species by the supplied energy. This partitioning in energy regimes agrees with the stable formation of a plasma plume at ICP powers set equal to or higher than 50 W. Plasma experiments in Sec. IV are conducted in the high-power regime.

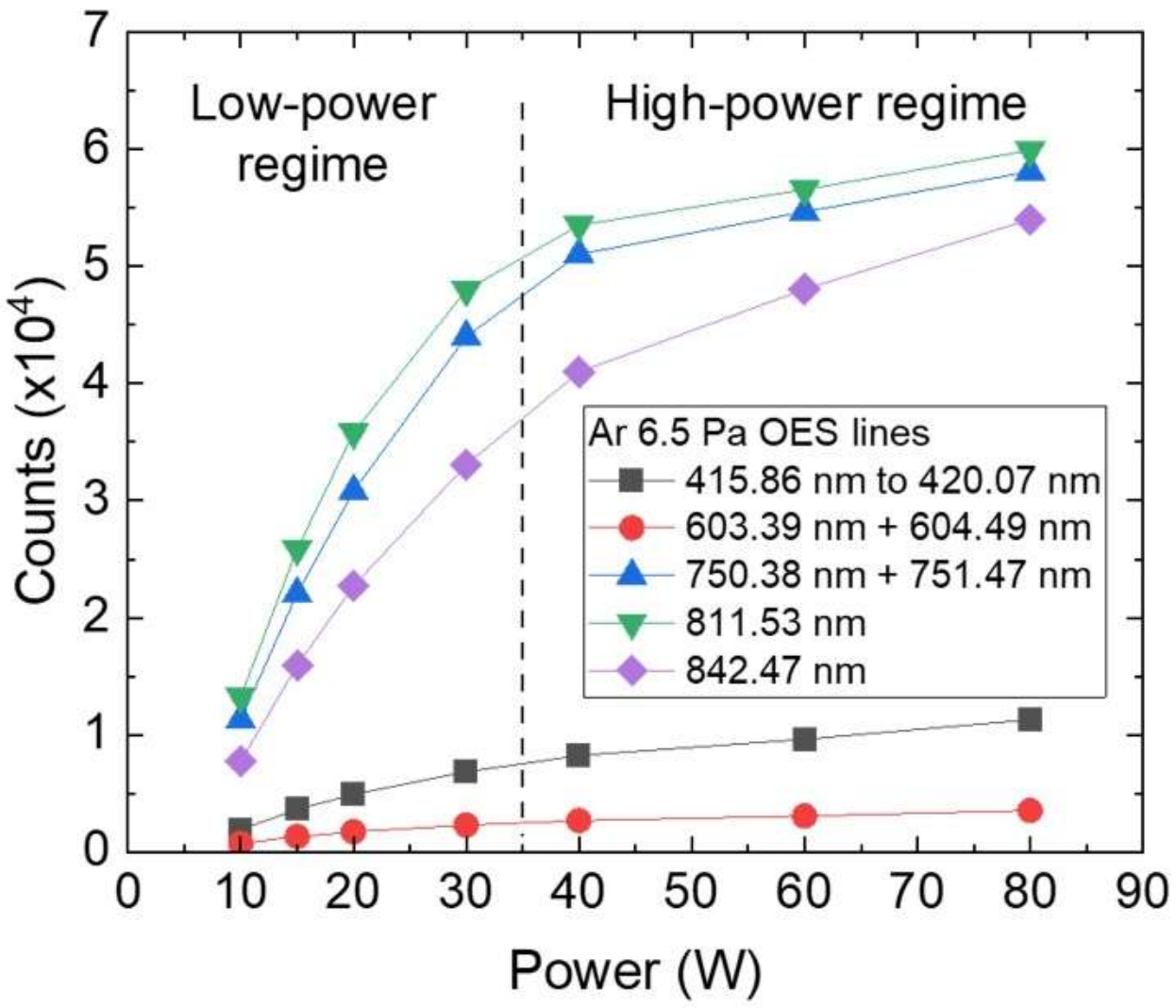

FIG. 10. Emission line (cumulative) intensities (OES) in Ar ICP discharge at 6.5 Pa within the RF power range between 10 W and 80 W corresponding to relevant optical transitions. The experimental uncertainty, corresponding to the OES signal variation during plasma measurement, is on the order of ±1 %. The vertical dashed line approximately separates the two power regimes.

In addition to the OES probe, the Si photodiode was used to measure the total light emission from the discharge with two main applications. First, the spatial brightness profile is a proxy for plasma density uniformity within the chamber. Second, time-dependent plasmas, such as pulsed discharges, produce light waveforms that are very useful for studying the instantaneous thermal flux impinging on the sample/substrate. An exemplary measurement has been discussed on Sec. IV.B.2.

### *2. Quadrupole mass spectrometry*

QMS is a well-established technique to study plasma chemistry in the discharge volume and reactor walls. These plasma-generated or thermally desorbed reactive species can appreciable contribute to the nanocalorimeter signal via exothermic/endothermic reactions at the sensor surface[57] and can be monitored with the proposed setup. In particular, a retractable stainless-steel tube with a 1 mm inner diameter extends to the center of the chamber from the rear flange (Fig. 1c, d). The capillary tube can be moved along the horizontal axis to collect plasma species with a spatial resolution. This channel extends to a vacuum-decoupled small cross chamber, pumped by a separate turbomolecular pumping system and connected to the main chamber via a copper gasket seal. There, the capillary tube delivers the plasma species probed in the main chamber through a small lateral orifice facing a vertically assembled residual gas analyzer (RGA). The RGA operated with a quadrupole charge/mass filter to detect chemical species via a Faraday cup within the range 1 u to 100 u. The mass spectra provided by RGA reveal the gas-phase composition of different plasma plume and/or chamber regions. The pressure in the small chamber is monitored with a separate Bayard-Alpert gauge.

Figure 11 compares the QMS spectra for Ar at 13.5 Pa with plasma ON and OFF. The QMS probe tip was 5 cm away from the chamber center. Water ($H_2O$) peak dominates plasma OFF residual gas spectrum as expected for an unbaked vacuum chamber, followed in decreasing intensity by peaks of OH, CO, $CO_2$, $H_2$, and other minor residuals. $Ar^+$ and $Ar^{2+}$ peaks dominate QMS when Ar gas is flowing into the chamber. The peak at 28 u has been mainly attributed to CO instead of $N_2$ because its ratio is significantly higher than the 4:1 relative to $O_2$ as expected for air. This fact, together with the abundant $CO_2$, reveals a high hydrocarbon concentration in the chamber. In comparison, an ultra-high vacuum system with baked walls is expected to exhibit much lower concentrations of water vapor and hydrocarbons, but the base pressure here is low enough to enable energy flux measurements using nanocalorimetry (Sec. IV).

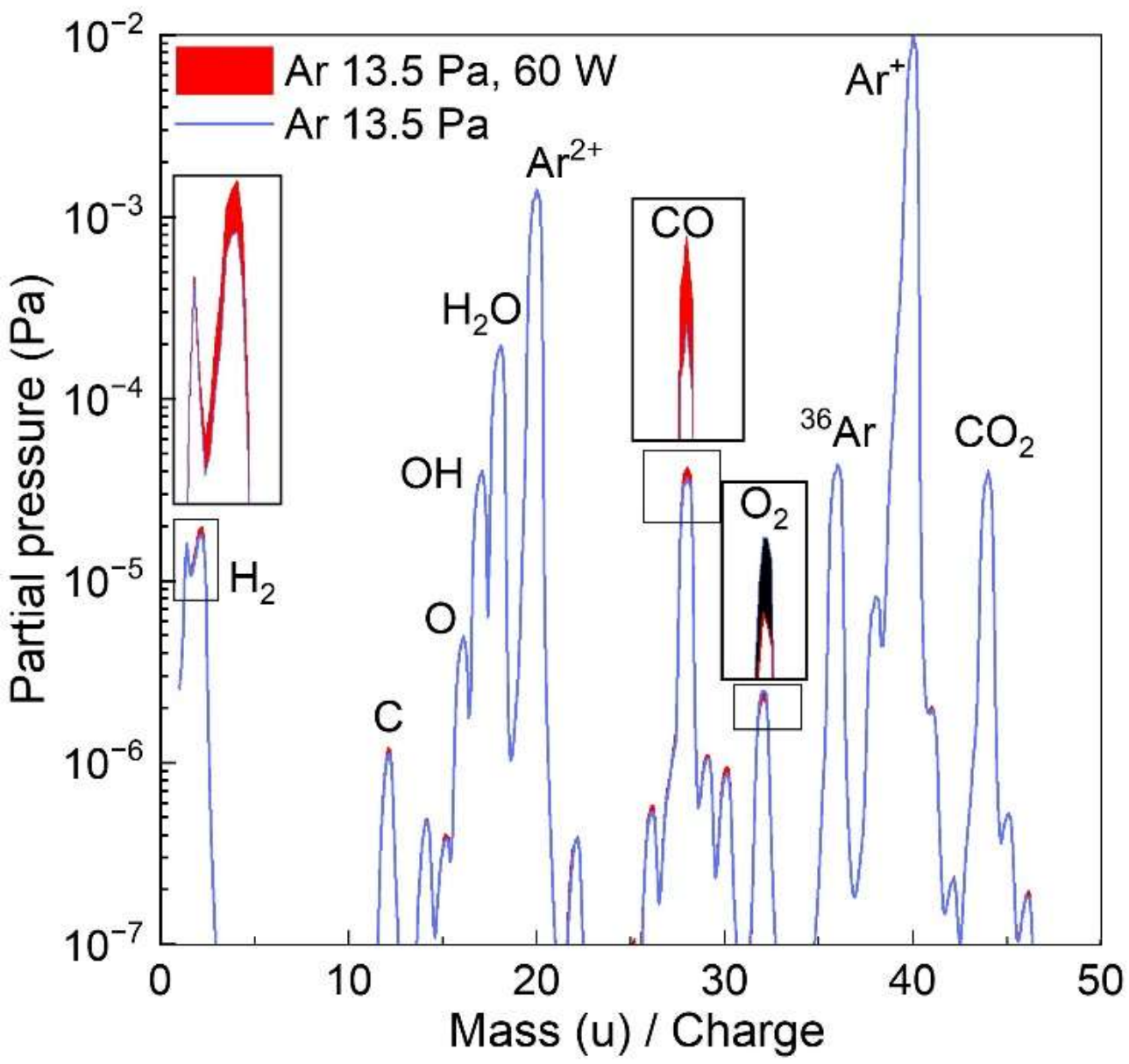

FIG. 11. Residual gas analyzer (RGA) spectra from Ar at 13.5 Pa (blue line) and Ar CCP discharge at 13.5 Pa and 60 W (red line and filling under curve). The main species are indicated, including $^{36}$Ar isotope. The inset images show zoom-ins of $H_2$, CO, and $O_2$ peaks.

As one can see, the concentrations of $H_2$ and CO increase upon plasma initiation, while the $O_2$ signal decreases. Figure 12 shows the corresponding detailed changes in the $H_2$, CO, and $O_2$ signals. The increase in $H_2$ and CO peak intensities during the plasma phase suggests plasma-induced dissociation of residual gases and wall degassing. Indeed, wall reactions induced by incident plasma species enhance gas impurities desorption. Another important source of the $H_2$ signal in the mass spectrum is the dissociation of the residual $H_2O$ vapor. In contrast, the $O_2$ case shows the opposite behavior: $O_2$ residual molecules are consumed when the plasma is active. This behavior is probably related to the formation of CO and $CO_2$ when plasma-activated oxygen species react with hydrocarbons on electrodes and chamber walls. Apparently, the latter process overwhelms the competing oxygen release from $H_2O$ dissociation.

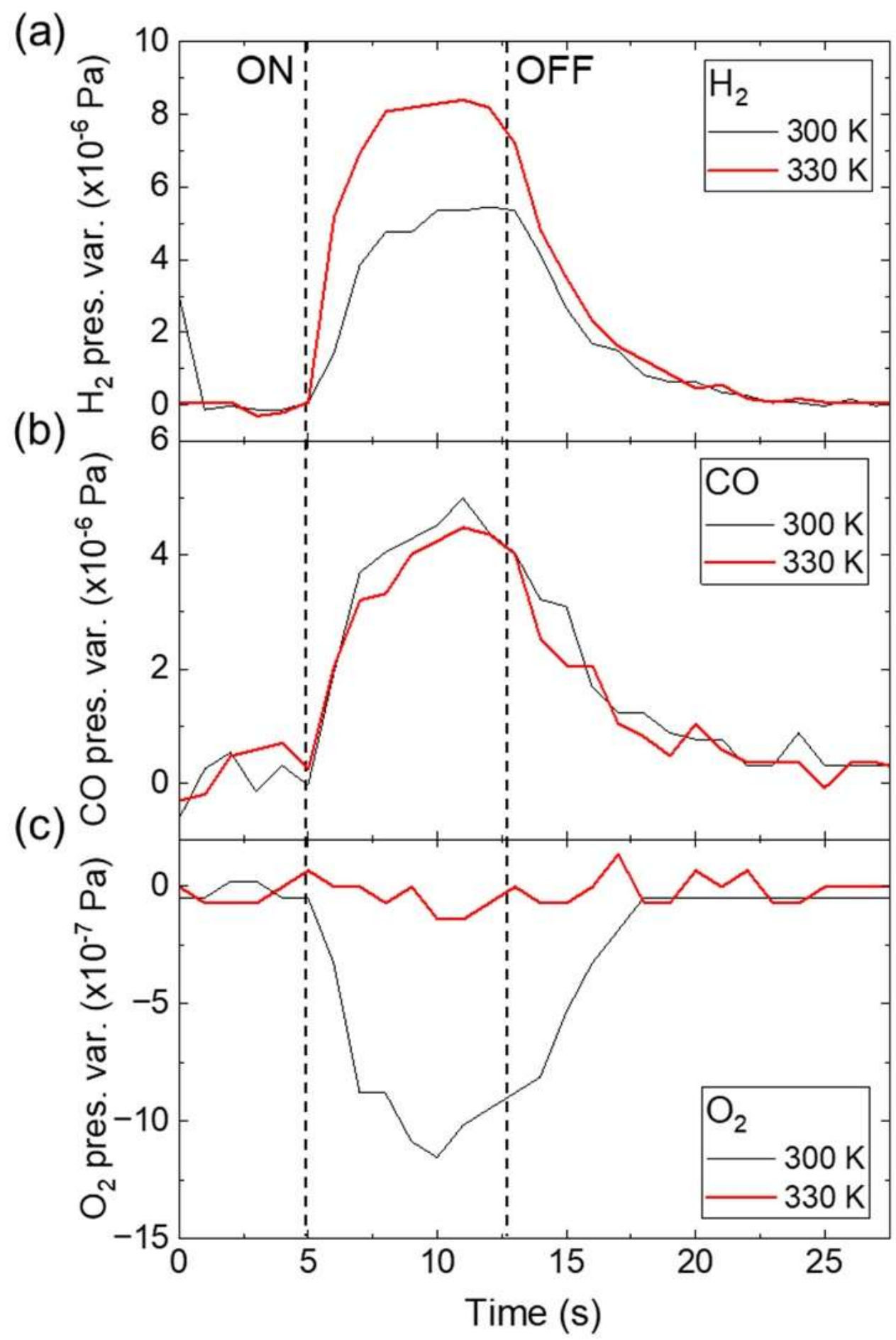

FIG. 12. Partial pressure variations with time for (a) hydrogen, (b) carbon monoxide, and (c) oxygen measured using QMS with sensor envelope at 300 K (black line) and 330 K (red line) during ON/OFF Ar CCP discharge at 13.5 Pa and 60 W. The plasma ON/OFF onsets are indicated with vertical dashed lines (time resolution: 1 s).

The same QMS experiments were repeated at different temperatures of the sensor envelope, when the reactor active parts become warmer after 30 min plasma-induced heating. The curves corresponding to the maximum temperature explored, 330 K, are also plotted in Figure 12. The $H_2$ and $O_2$ channels show higher partial pressure presumably due to an increase in the water desorption rate from warmed-up walls. The curves tend to reproduce the initial profiles as the system cools to room temperature.

The temporal evolutions of QMS signals were measured with a resolution of one second. Fig. 12 shows that concentration variations occur within a few seconds of plasma ignition and extinction. This characteristic time represents the average residence time of the gas elements in the chamber and will be compared with the stabilization time of the nanocalorimeter's temperature curves in Sec. IV.B.2.

## *C. Thermocouple probe for plasma temperature profiling*

The Langmuir probe could be replaced by a local-temperature probe to map the radial temperature profile in the chamber[58]. A small, bare thermocouple was mounted at the tip of a wire protected by a thin ceramic tube (2.75 mm outer diameter). Similar to the Langmuir probe setup, the thermocouple could be moved along the chamber axis with minimal disturbance to the discharge region. Temperature measurements were intended to characterize the plasma temperature profile relevant to nanocalorimetry experiments. The K-type thermocouple used in these experiments has a sensitivity of 25 K/mV.

The temperature profiles of CCP discharges in Ar and He have been recorded using the thermocouple at five different locations along the horizontal axis. Figure 13 shows the temperature responses to Ar CCP discharge at 13.5 Pa and 60 W for approximately 90 s. Only a few degrees of temperature increase were measured at locations outside the CCP cell, whereas the temperature increased by around 20 K in the cell center. Note that the thermocouple's heat capacity is much higher than that of the nanocalorimeter, which slows down the heating/cooling rates when plasma is switched ON/OFF (see Sec. IV).

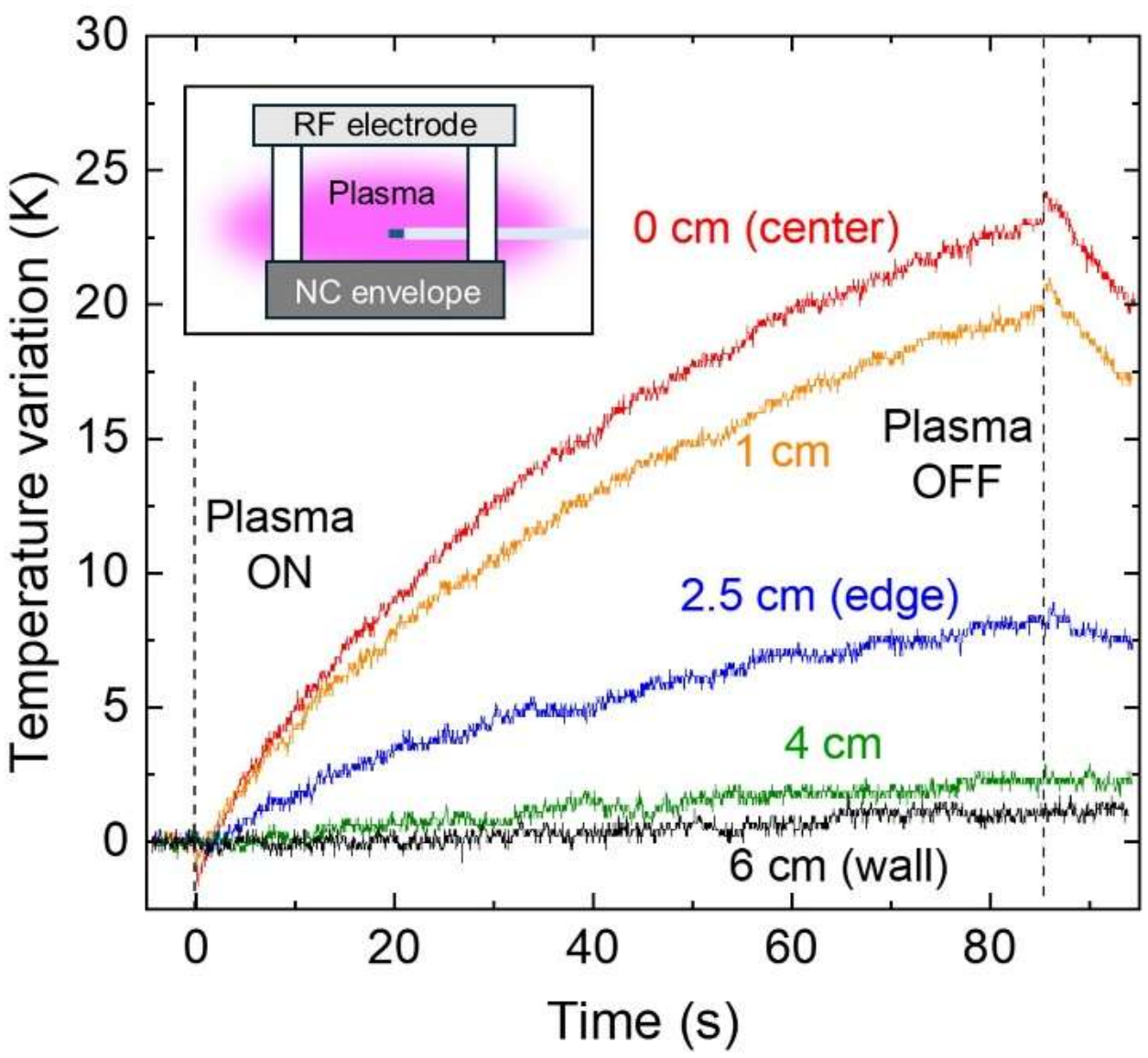

FIG. 13. Evolution of temperature with time at different distances of the thermocouple tip from the center for an Ar CCP discharge at 13.5 Pa and 60 W. Inset: Sketch of the thermocouple probe positioned at the center of the CCP cell, between the RF-biased electrode and nanocalorimeter (NC) envelope.

Temperature profiles for Ar and He CCP discharges after approximately 90 s of plasma exposure are depicted in Figure 14. It is worth noting that all the thermocouple measurements discussed here were performed in the glow region of the discharge. Temperatures measured in the dark regions, reported in the *Supplementary Material*, were systematically lower, as expected for lower plasma densities.

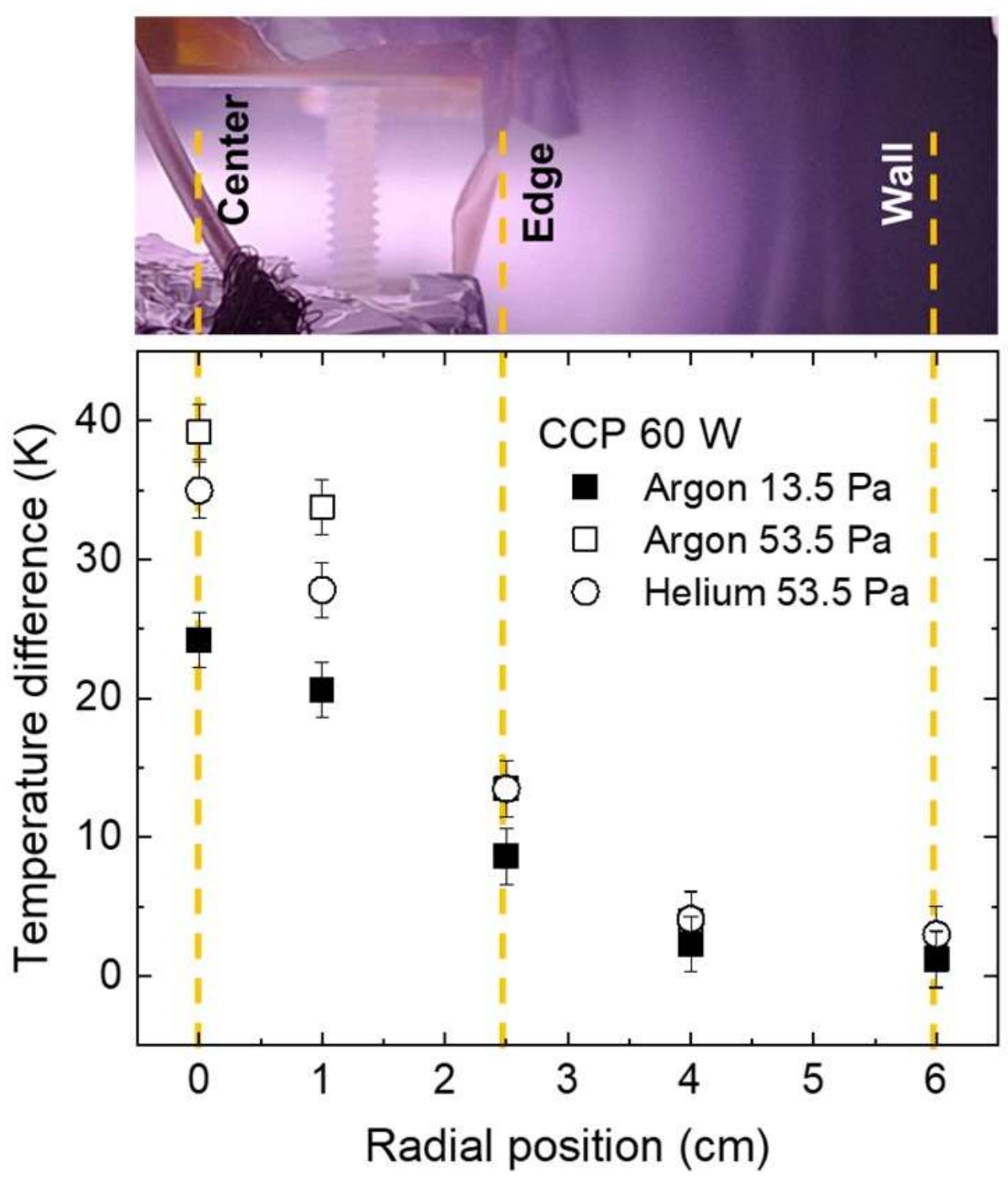

FIG. 14. Recorded temperature in Ar and He CCP discharges at 60 W RF as a function of the thermocouple's radial position. Top image shows the corresponding zones within the CCP discharges. The vertical error bars, proportional to the temperature noise, are approximately ±2 K.

# IV. NANOCALORIMETRY APPLICATIONS EXAMPLES

This section presents two applications of plasma nanocalorimetry: discrimination between ion- and electron-induced thermal effects via DC biasing of a sensor immersed in an ICP discharge, and detection of temperature transients during pulsed discharge monitoring (CCP configuration). In a previous article, the metrology of reactive plasma radicals via differential nanocalorimetry was reported for a hydrogen ICP RF discharge[40]. In that case, gold catalyst and inert alumina coatings were used as active and reference surfaces, respectively, to quantify hydrogen radical fluxes. The heat released when H atoms recombine on a gold surface to form $H_2$ molecules was used as a thermal signature for radical detection. The results were corroborated by independent measurements of hydrogen radical density using optical actinometry.

In this report, a bare $SiN_x$ membrane was used, which serves as a chemically inert surface for the gases employed (Ar, $O_2$). Since no chemical reactions are expected on the sensor's surface for such a configuration, the measured temperatures correspond to the gas temperature near the sensor plus smaller heating contributions from charged particles, energetic neutrals, and photons[21].

## A. Decoupling of plasma density and ion energy – ICP configuration

An important application of plasma material processing is etching via reactive and/or energetic ions[4,59]. A plasma nanocalorimetry setup can be designed to study the processes involving energetic ion bombardment. Indeed, the energy of impinging ions can be altered by conveniently setting a sensor potential. In this way, the thermal effect of ions can be discriminated against other contributions (electrons, neutrals, and photons). The sections below provide detailed explanations of the estimation of energy fluxes from an ICP discharge, performed under conditions comparable to those in experiments reported earlier in ref.[21]. The present report therefore complements the initial research on the nanocalorimetry responses to plasma heat fluxes.

### 1. Time domains

Figure 15 shows the temperature of the nanocalorimeter sensor mounted on a grounded electrode and exposed to an Ar ICP discharge at 6.5 Pa and 80 W. Two time domains can be observed in a sensor response to plasma ignition[21,40]: (1) rapid sub-second temperature jump triggered by an onset of plasma exposure (shutter opens), followed by (2) slow (hundreds of seconds) temperature increase. The short sensor response time is due to the small thermal mass and the low thermal conductivity of the Pt/$SiN_x$ membrane stack. The slow temperature increase is due to the gradual plasma-induced warming of the chip's Si frame and Al envelope, which have a significantly larger thermal mass than the nanocalorimeter. This progressive heating continues until saturation is reached (not shown here), indicating that the incoming and outgoing heat fluxes in Eq. (1) are balanced. The thermal losses of the entire sensor assembly are mainly due to heat conduction to the gas environment and, eventually, to the chamber wall through the metal assembly parts.

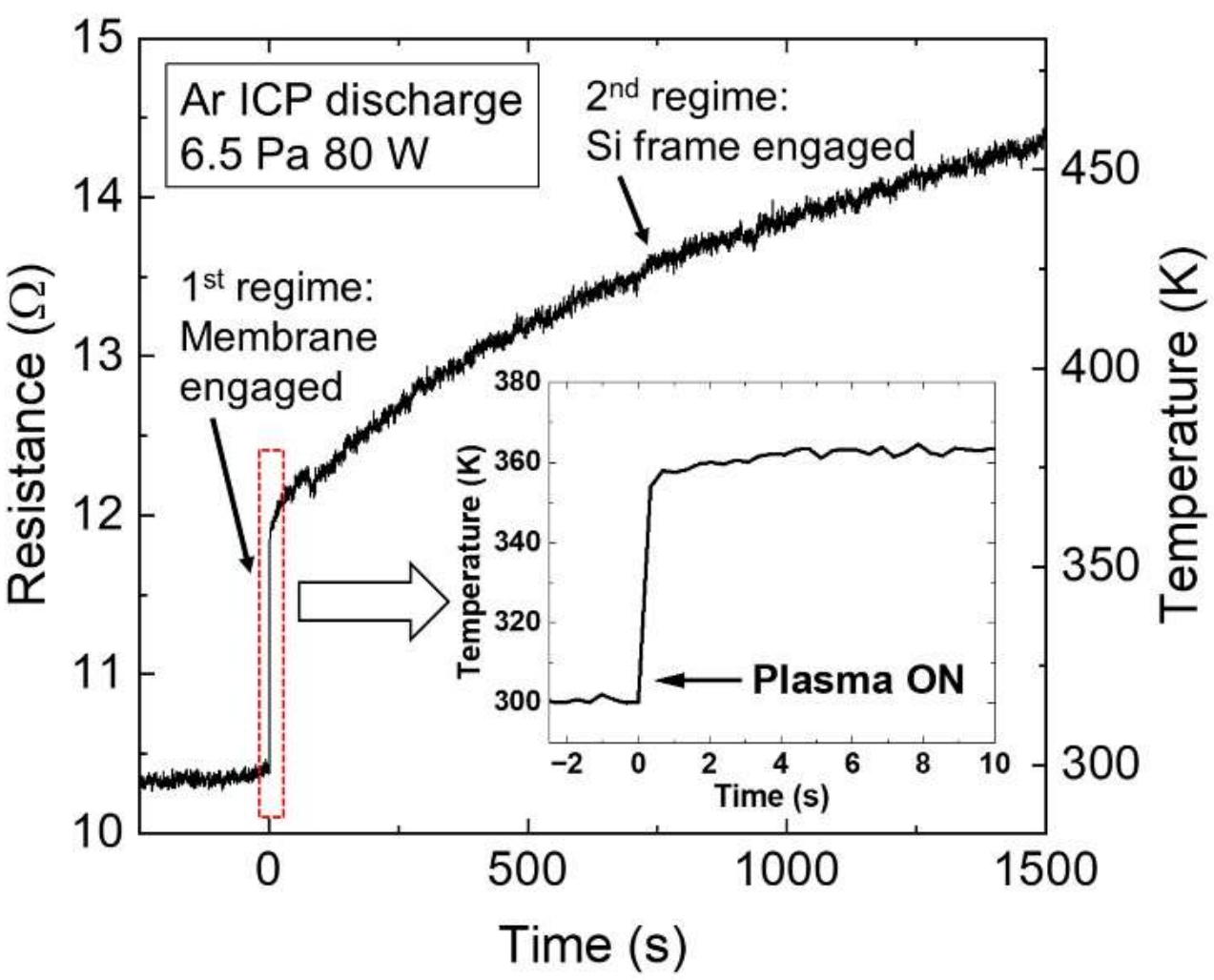

FIG. 15. Variation of Pt strip resistance and temperature of grounded nanocalorimeter exposed to an Ar ICP discharge at 6.5 Pa and 80 W RF. Two time-domains/regimes for sensor heating are indicated. Inset: Zoomed initial part of the curve (red dotted rectangle) showing a few seconds after plasma onset (shutter open).

The incoming energy flux can be evaluated by applying Eq. (1) to the fast transient at the onset of plasma ignition[7], because the outgoing energy fluxes can be approximated as zero at this moment ($P_{\mathrm{out}}$ = 0). The estimated incoming heat flux through the Al slit to sensor is approximately 1.5 mW, which corresponds to 470 W/m$^2$ in energy flux density from plasma.

## 2. *Sensor biasing*

Here, variations in energy flux from an Ar plasma with changes in sensor potential are evaluated. RFEA measurements were performed beforehand to characterize the energy distribution of incoming ions and to define the major peak energy. As shown in Figure 16a, the average peak energy increases linearly with DC bias applied to the Pt strip. As reported elsewhere, the IEDF becomes a monomodal distribution with a smaller FWHM as the sensor potential increases[21]. The kinetic energy peak of plasma ions accelerated across the sheath is[38]:

$$E_{\mathrm{i}} = e\left(V_{\mathrm{p}} - V_{\mathrm{bias}}\right) \tag{2}$$

where $e$ is the elementary charge ($1.602\times10^{-19}$ C), $V_{\mathrm{p}}$ is the plasma potential, and $V_{\mathrm{bias}}$ is the applied bias. Only singly ionized Ar species were considered, given the significantly lower density of doubly ionized species, as it is assumed for the moderate values of electron temperature (Sec. III.A.1). For instance, the peak energy on the grounded sensor was around 30 eV, which corresponds to the plasma potential of 30 V measured with the Langmuir probe.

Along with peak energy, ion flux also increases with sensor potential (Figure 16b). Ion current densities were collected using both RFEA and the nanocalorimeter itself. The latter, which were obtained using a current meter connected between the DC power supply and the nanocalorimeter (Fig. 4), show systematically lower values than the RFEA readings. Such a discrepancy is probably due to the finite electrical conductivity of the sensor's non-stoichiometric $SiN_x$ membrane. Besides, it may eventually result in the buildup of a retarding potential, which could affect the ion's impinging energy of the ions and the sensor's response time. Another factor likely contributing to the systematic difference between RFEA and nanocalorimeter current densities is the approximation of the effective plasma exposure area to the slit opening area (0.7 mm × 4.6 mm).

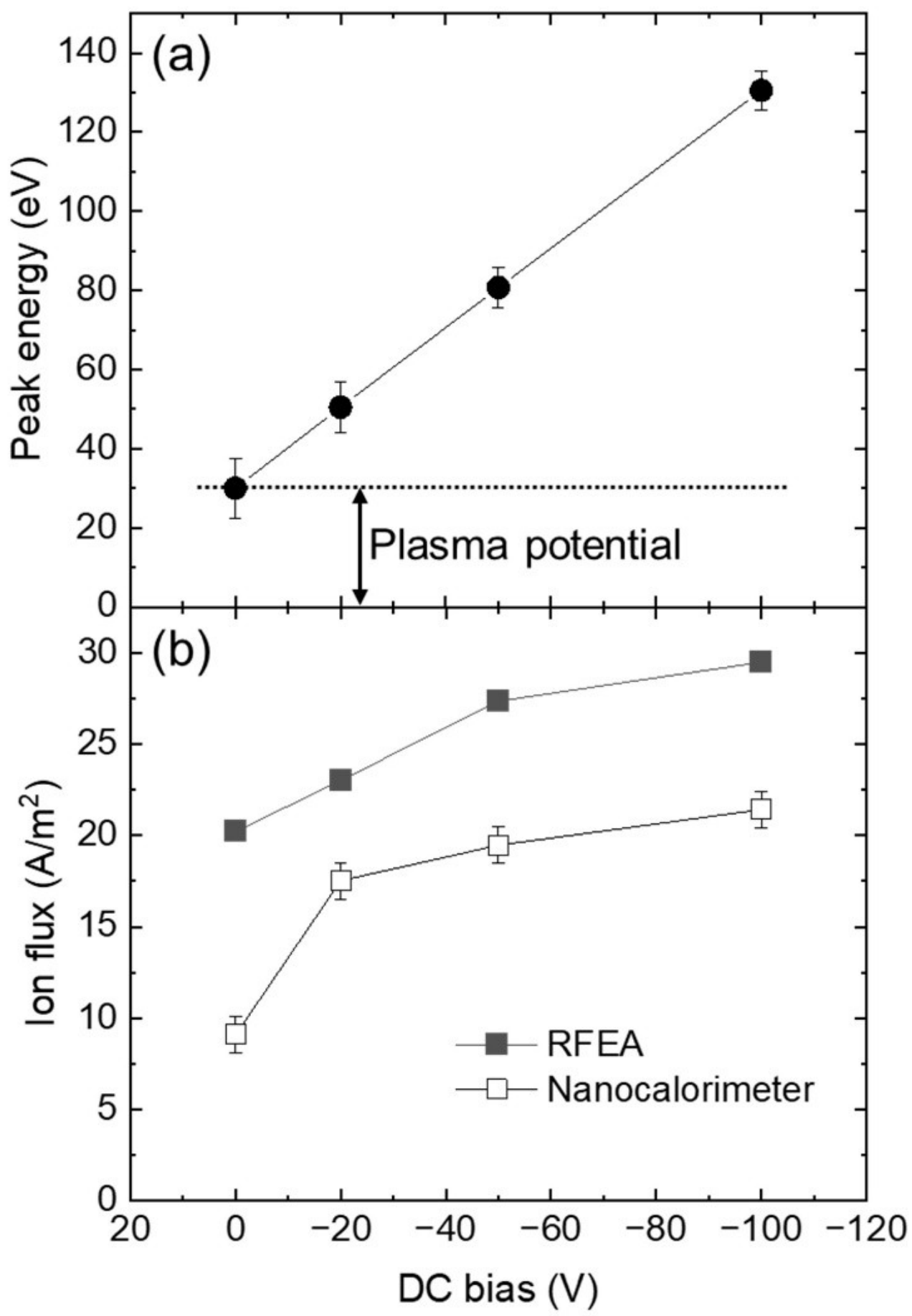

FIG. 16. (a) Linear variation of ion energy distribution function (IEDF) average peak energy in Ar ICP discharges held at 6.5 Pa and 80 W, measured with retarding field energy analyzer (RFEA), with external DC bias applied to the analyzer. The vertical error bars account for the full width half maximum (FWHM) of each ion energy distribution function (IEDF). The dashed line labeled as "plasma potential" was measured with a Langmuir probe. (b) Variations of ion

flux measured using the RFEA and nanocalorimeter are compared. The statistical uncertainties for RFEA- and nanocalorimeter-measured ion fluxes are limited to ±0.5 A/m$^2$ and ±1 A/m$^2$, respectively.

Figure 17a shows the temperature measured by an electrically biased sensor, exposed to Ar ICP discharges at 6.5 Pa and 80 W. Sensor temperatures exhibit a stepwise profile as the DC bias is incrementally changed in 120 s voltage steps of ±10 V up to a voltage peak of -30 V. In contrast with ref. [21], further voltages were not explored here to minimize the risk of sensor break. An increase or decrease in DC bias leads to a corresponding decrease or increase in temperature due to the variations of $Ar^+$ ion flux and mean ion energy. A second temperature curve, corresponding to sensor heating at 0 V (grounded) throughout the process (another experiment) is shown for comparison. The small offset in resistance observed between the two curves is due to a temperature difference at the initial plasma-ON conditions.

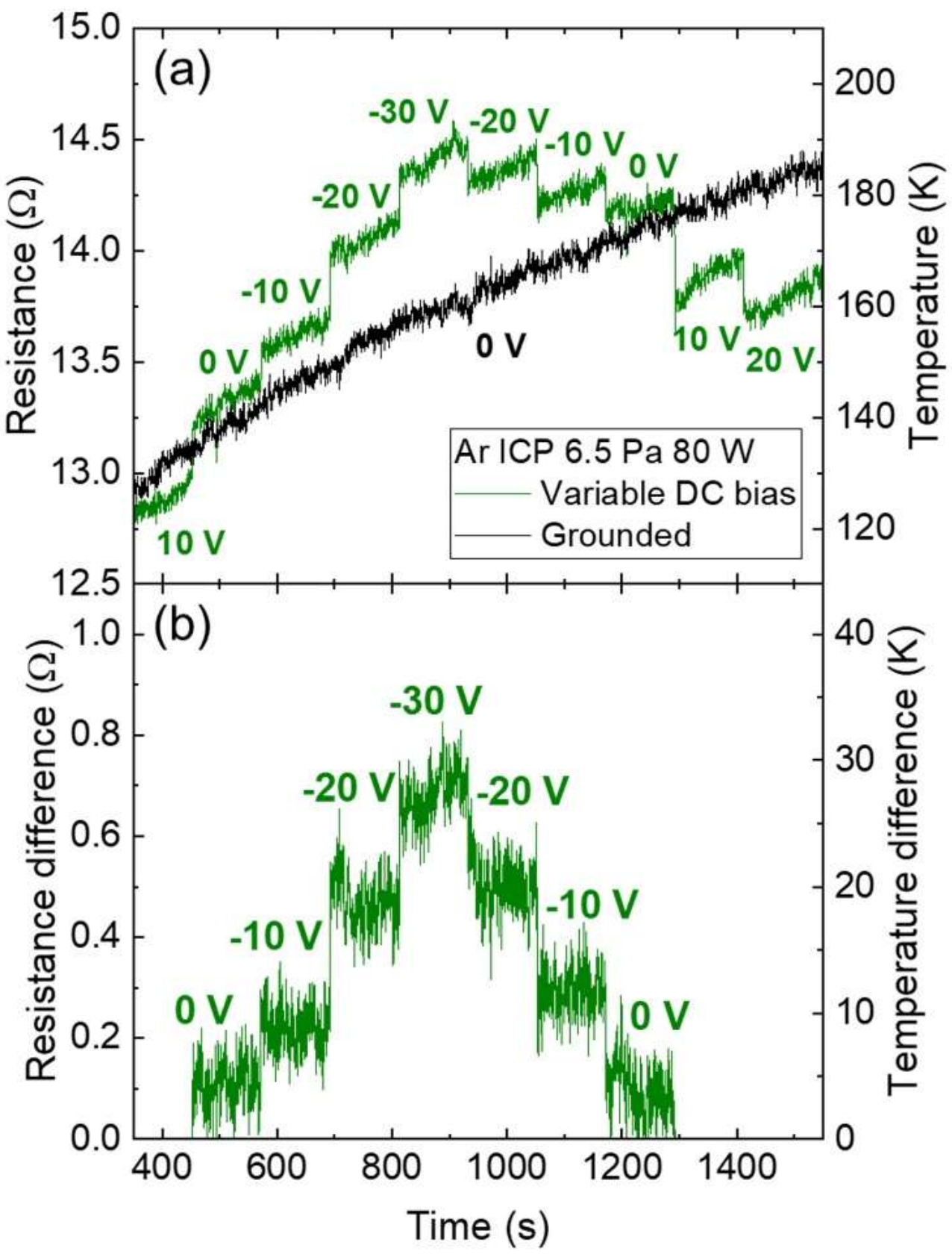

FIG. 17. (a) Green line: Stepwise evolutions of nanocalorimeter resistance and corresponding calculated temperature as sensor's DC bias decreases from 10 V to -30 V and increases from -30 V to 20 V in an Ar ICP discharge at 6.5 Pa and 80 W. Black line: Nanocalorimeter readings for same plasma collected with a grounded sensor (0 V) (b) Staircase-shaped curve resulted from subtraction of the curves above to compensate for temperature drifts in sensor readings.

Figure 17b shows the staircase-shaped temperature curve after subtraction of the drift temperature measured when the sensor is grounded. The temperature steps observed in the subtracted curve are attributed exclusively to sensor heating by plasma ions, which collide with the sensor surface with a kinetic energy $E_{\mathrm{i}}$ [Eq. (2)] if we ignore the low-energy tail in IEDF curves (Sec. III.A.2). Moreover, the resulting curve proves that sensor heating via modulation of the ion bombardment is a reversible process. The temperature steps are limited to $\Delta T \approx 10$ K (with $\Delta t \approx 50$ ms), which corresponds to an ion energy flux of around 100 W/m$^2$ per step according to Eq. (1) and considering the effective heat capacity and opening slit area. Such a value is a small fraction of the total energy flux from plasma (470 W/m$^2$), which also includes contributions from electrons, photons, and energetic neutrals in addition to ions. No significant ion-induced sputtering was observed in this experiment, which corroborates with Transport and Range of Ion in Matter (TRIM) simulation[60] of an etching rate to be 0.01 nm/s for the $SiN_x$ membrane, assuming maximal ion kinetic energy to be 60 eV for the -30 V bias (see Figure 16a).

Positive DC bias voltages have also been considered here to modulate electron-induced thermal fluxes. However, high electron currents occur already at small positive biases, and the accompanying Joule heating damages the fragile $SiN_x$ membrane. It is worth noting that the effect of the particles' kinetic energy is not discriminated against the increase in the ion flux in the reported measurements because both variables respond simultaneously to DC bias variations.

## ***B. Study of pulsed plasma transients – CCP configuration***

The use of pulsed plasmas is very common in semiconductor manufacturing to control heating rates of substrates and physical properties/quality of the deposited coatings[61-63]. Here, we demonstrate the nanocalorimeter's ability to track substrate temperature during pulsed discharges. In particular, this technique has promising diagnostic applications in high-power pulsed discharges with typical duty cycles below 1 %, such as high-power impulse magnetron sputtering (HiPIMS) discharges[49,63,64].

### *1. Periodic pulses*

Figure 18 shows temperature oscillations measured in pulsed Ar CCP discharges held at 13.5 Pa with RF powers of 40 W and 60 W. The pulse frequency and duty cycle were 2 Hz and 50 %, respectively, ensuring complete plasma decay between pulses. The sensor's voltage drop was recorded with an oscilloscope operating at 2 GSa/s and 200 MHz bandwidth. The minimum recorded temperatures at the sensor are near room temperature, while the maximum values are around 320 K for 40 W and 330 K for 60 W. The related energy fluxes were estimated by using the plasma ON/OFF onsets according to Eq. (1) as described above. An average value of 250 W/m$^2$ over $\Delta T$ = 20 K is measured at 40 W RF power, and it increases to 400 W/m$^2$ over $\Delta T$ = 30 K for 60 W. The latter value is equivalent to 1.2 mW through the Al envelope slit opening.

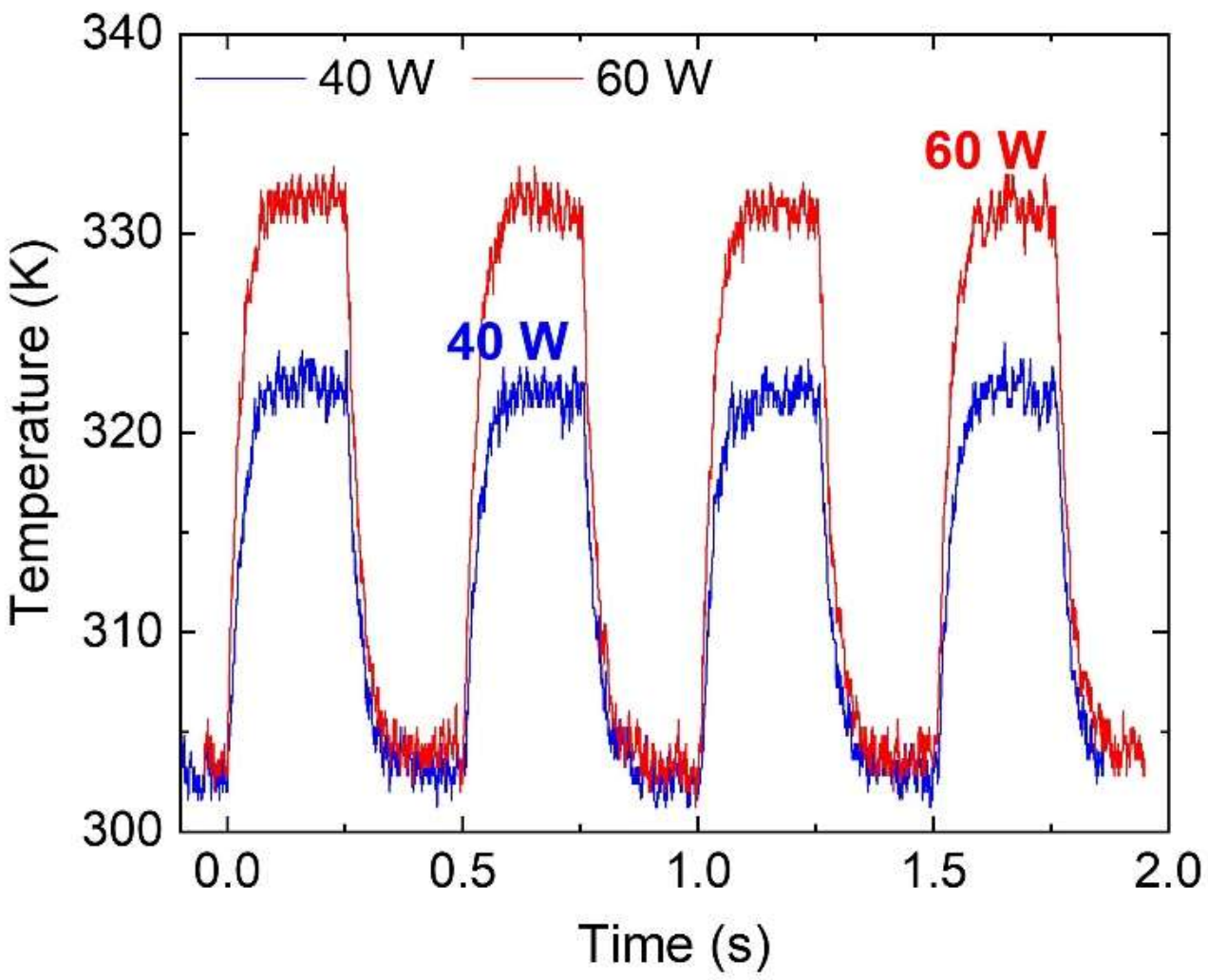

FIG. 18. Temperature waveforms measured by nanocalorimeter sensor exposed to CCP discharges of Ar at 13.5 Pa, 40 W (blue curve) and 60 W (red curve), driven with a pulsed RF power supply operating at 2 Hz, 50% duty cycle.

The energy fluxes recorded by the nanocalorimeter are proportional to the supplied RF power, which in turn is proportional to the plasma density (see Figure 6). Note that the measured fluxes include heating contributions not only from neutral species but also from charged particles: the grounded sensor is exposed to $Ar^+$ ions with a peak kinetic energy corresponding to the plasma potential, which is around 30 V, similar to our ICP discharges. Therefore, ion etching of the $SiN_x$ membrane is inefficient since the ion kinetic energy at the sensor is near or below the sputtering threshold for $Ar^+$ ions[65,66].

### *2. Response time*

Nanocalorimeters can monitor thermal fluxes in pulsed discharges with time resolution down to less than a millisecond for optimized thickness and lateral dimensions[27]. In this experiment, the silicon photodiode (Sec. III.B.1) has been used to record the total light intensity emitted by pulsed discharges synchronously with sensor temperature. Figure 19 depicts the responses of nanocalorimeter temperature and the photodiode signal to a square-wave modulated CCP discharge of Ar at 13.5 Pa and 40 W. Again, the nanocalorimeter signal shows a sub-second response followed by a slower temperature drift along the pulse due to heat exchange with the environment (Ar gas and walls). Optical emission shows a sub-millisecond ignition time. In contrast, plasma afterglow time is in the order of 15 ms.

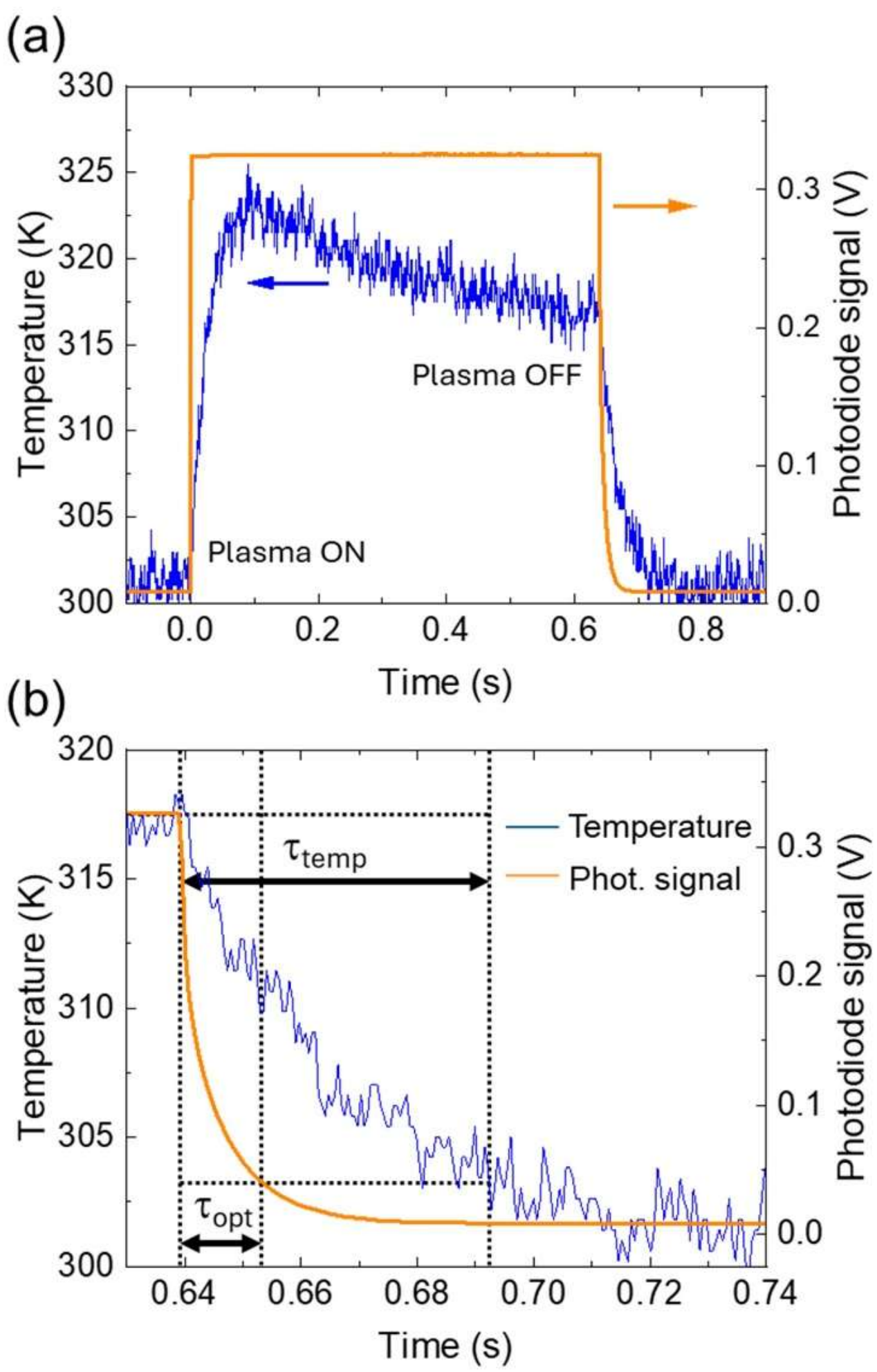

FIG. 19. (a) Response of sensor's temperature (blue curve) and photodiode signal (orange curve) to a pulse of square-wave modulated RF CCP discharge in Ar at 13.5 Pa and 40 W. (b) Zoom in the region of plasma OFF, indicating the decay times for thermal signal, $\tau_{\mathrm{temp}}$, and optical signal, $\tau_{\mathrm{opt}}$.

Here, response time is defined as the time to reach 90 % of the full signal value for both sensors. Rise and decay times for the nanocalorimeter sensor were around 50 ms, which were of the same order of magnitude as the characteristic time for heat propagation in the gas through collisions among heavy species[40]. The QMS signal variation is of the order of one second, as shown in Sec. III.B.2, thereby showing that the timescale for thermal stabilization is shorter than the gas residence time by one order of magnitude. The optical signal evolution was much faster than the typical thermal processes, which is attributed to the relatively fast electron-controlled gas ionization/relaxation events in the plasma state. The asymmetry observed in the optical ignition and afterglow times is common and is related to the fast electron multiplication during the ignition phase and to slower ion neutralization and quenching of Ar metastable atoms when the plasma is OFF[67].

## V. CONCLUSIONS AND OUTLOOK

An experimental setup for evaluating nanocalorimetry metrology for plasma thermal diagnostics has been described. The microfabricated nanocalorimetric sensors demonstrated capabilities to quantify reactive radicals, energetic ion fluxes, and other plasma species. In our setup, the nanocalorimeter sensor is employed in combination with traditional plasma diagnostics tools, such as Langmuir probe and RFEA, used to pre-characterize discharges (Table III). Plasma potential, ion flux, and IEDF obtained by these tools have been correlated to sensor heating induced by plasma species. OES and time-resolved emission intensity measurements were used to analyze plasma chemistry and power coupling to the ionized gas. QMS provides insight into gas contamination, revealing how purity affects the discharge's thermal behavior. Finally, the data from the thermocouple probe were helpful in evaluating the spatial distribution of the heat flux using spatially resolved measurements. The available collection of plasma diagnostics and adaptability to various plasma sources relevant to material processing, such as magnetron sputtering cathodes, demonstrates the versatility of the plasma setup for the nanocalorimetry study.

TABLE III. Relations between plasma diagnostic techniques and nanocalorimetry measurements of plasma energy fluxes. RFEA: Retarding field energy analyzer; OES: Optical emission spectroscopy; QMS: Quadrupole mass spectrometry; EEDF: Electron energy distribution function; IEDF: Ion energy distribution function.

| **Characterization** | **Technique** | **Output** | **Input for nanocalorimetry** |
|---|---|---|---|
| Electrical | Langmuir probe | EEDF and plasma parameters | Basic plasma pre-characterization, plasma potential and estimation of the ion impinging energy and related heat flux |
| | RFEA | IEDF and ion flux | Direct measurements of the energy fluxes by ions and electrons |
| Opto-spectrometric | OES/photodiode | Optically active species and plasma uniformity | Power regime and response time, VUV-vis-IR contribution to the heat flux |
| | QMS | Plasma chemistry / wall reactions | Potential contribution of impurities and reactive plasma species to energy flux |
| Thermal | Thermocouple probe | Gas temperature | Thermal status of plasma volume |

Nanocalorimeter sensors are suitable for tracking plasma-induced heating processes on timescales of several tens of milliseconds due to their $1.5\times10^{-6}$ J/K thermal mass. Such a performance is a promising addition to modern plasma calorimetry because, as mentioned above, plasma thermal probes have usually operated with sensitivities of 0.01 J/K or higher. Nanocalorimeters operated with optimized electronic parameters demonstrate sensitivity down to fractions of a Kelvin[40], corresponding to a heat flux below 10 W/m$^2$, unaffected by plasma operation. The readout, consisting of sensor temperature, combined with short response times that are competitive with those of the fastest heat flux microsensors (see Table I), makes nanocalorimeters unique tools for real-time thermal analysis of plasma discharges. In addition, selectivity to specific plasma species can be enhanced when using a catalytic coating or by applying a DC bias to retard/accelerate ions or electrons. However, sensor potential should not be restricted to DC biasing. AC bias excitation (not used in this study) can also be applied to induce plasma-assisted thermal processes, which can be analyzed in the frequency domain[68]. Finally, further size reduction and arrangement of multiple sensors in a compact multiprobe array for multivariate diagnostics can be envisioned as the next refinement in plasma calorimeter technology.

An important issue in the performance of microfabricated nanocalorimeters is their fragility: they exhibit limited lifetimes under energetic and/or reactive ion etching, as well as during film formation in plasma-assisted deposition. To avoid early degradation, nanocalorimeters should be composed of materials compatible with the plasma atmospheres typical in specific applications. For instance, thicker membranes with inert coatings will be preferable to slow down the degradation of nanocalorimeters exposed to harsh environments. In our plasma setup, using QMS will help assess the atmosphere's chemistry before any calorimetry test. The thermal profile within the chamber can be tracked simultaneously with the thermocouple probe to spot the hot regions. The sensor status or mass load can be controlled periodically by in situ monitoring of heat capacity, and this readout can be verified by QCM measurements using the RFEA's built-in crystal monitor. The same QCM, if coated with a selected material such as carbon, is suited to measure physical/chemical etching rates combined with nanocalorimetry. Also, the optical properties of the sensor's top layer must be considered, as they are crucial to its emissivity, especially at high temperatures. The heat absorption from incident thermal radiation, mostly IR photons from hot surfaces, can be rapidly quantified by our sensor.

The technological challenges mentioned above can be addressed with the plasma monitor suite presented here, and they are expected to motivate the design of robust nanocalorimeters for their successful implementation in cold plasma metrology.

## SUPPLEMENTARY MATERIAL

A separate PDF file provides: (i) resistance-temperature calibration data and heat capacity values for a nanocalorimeter, (ii) a discussion on the collisional regime of ICP discharge sheaths, and (iii) an insight on thermocouple measurements in glow and dark plasma regions.

## ACKNOWLEDGEMENTS

This work was funded by the CHIPS Metrology Program, part of CHIPS for America, National Institute of Standards and Technology, U.S. Department of Commerce. CHIPS for America has financially supported this work through the “Nanocalorimetry for Semiconductors and Semiconductor Process Metrology” project. Certain commercial equipment and software are identified in this paper to foster understanding. Such identification does not imply recommendation or endorsement by the National Institute of Standards and Technology, nor does it imply that the materials or equipment identified are necessarily the best available for the purpose. The sensors were fabricated at the NIST Center for Nanoscale Science and Technology (CNST). The authors are thankful to Dr. Dylan Kirsch, Dr. James E. Maslar, Dr. Evan Groopman, and Dr. Mark McLean (all at NIST) for their careful reading of the manuscript and valuable suggestions. The feedback provided by external reviewers, which helped improve the article, is gratefully acknowledged.

## AUTHOR DECLARATIONS

### Conflicts of Interest

The authors have no conflicts to disclose.

### Author Contributions

**Carles Corbella:** Conceptualization (equal); Methodology (equal); Data curation (lead); Formal analysis (lead); Investigation (lead); Writing – original draft (lead). **Feng Yi:** Funding acquisition (lead); Project administration (equal); Resources (equal); Writing – review & editing (supporting). **Andrei Kolmakov:** Conceptualization (equal); Methodology (equal); Project administration (lead); Resources (equal); Software (supporting); Supervision (lead); Writing – review & editing (equal).

## DATA AVAILABILITY

The data that support the findings of this study are available from the corresponding author upon reasonable request.

## Supplementary Material for:

# Experimental setup for testing nanocalorimeter sensors as a plasma diagnostics tool

Carles Corbella[1,2], Feng Yi[1], Andrei Kolmakov[3]

[1] Materials Measurement Science Division, MML, NIST, Gaithersburg, MD 20899, USA

[2] Department of Chemistry & Biochemistry, University of Maryland, College Park, MD 20742, USA

[3] Nanoscale Device Characterization Division, PML, NIST, Gaithersburg, MD 20899, USA

## *A. Resistance-temperature calibration and heat capacity data*

Each nanocalorimeter was calibrated over two temperature ranges. First range comprised from room temperature to 360 K via external heating using an oven incubator. Second range was based on thermal emission from 600 K up to 900 K due to current self-heating measured by optical pyrometry. Figure S1 shows an example of calibration curve. The sensor's heat capacity was determined by applying electrical power variations and measuring the resulting temperature derivatives. The evolution of heat capacity values from room temperature through 900 K is shown in Figure S2.

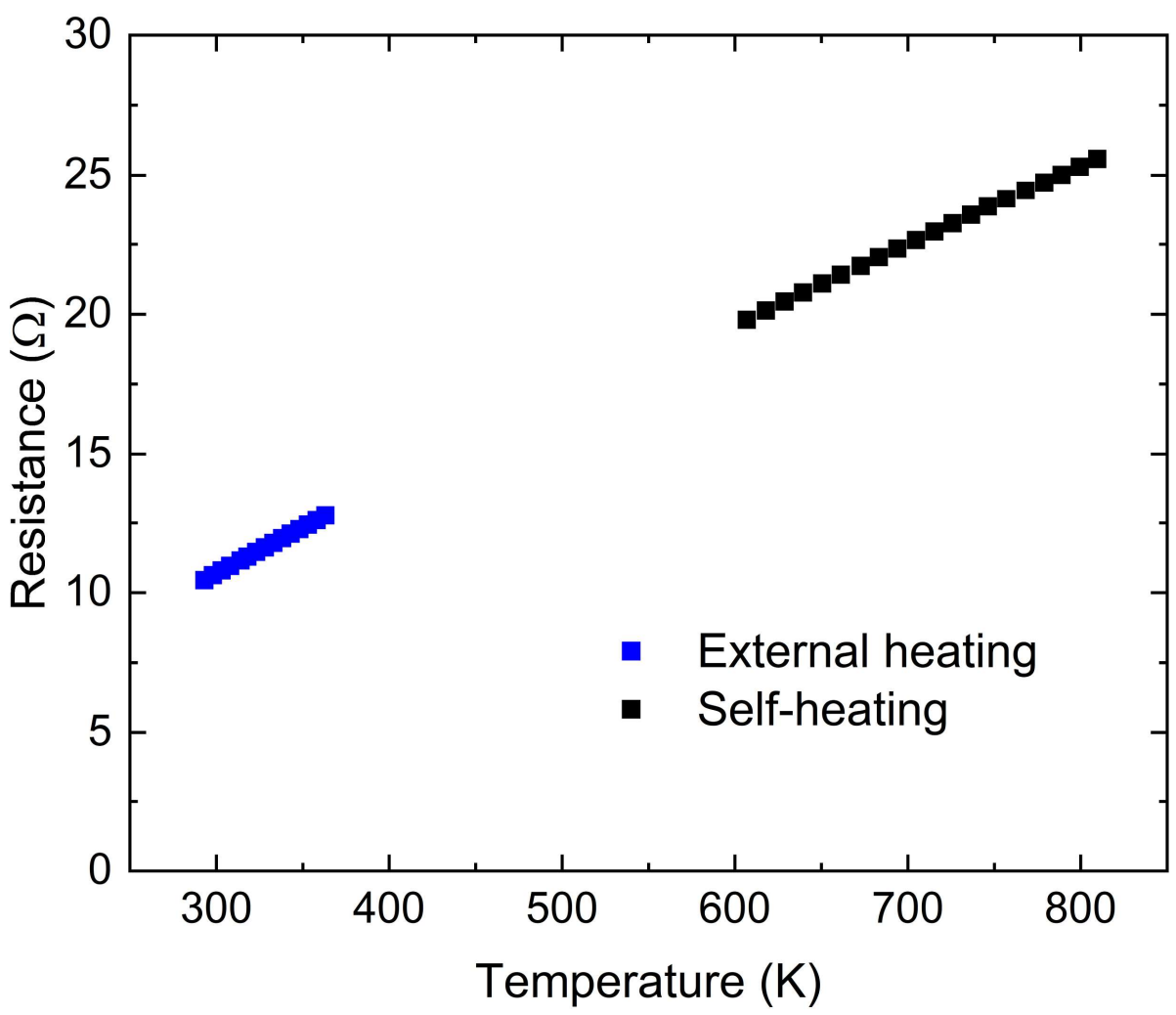

FIG. S1. Calibration measurements of a nanocalorimeter at low- and high-temperature regimes. The average temperature coefficient of resistance (TCR) is around 0.03 Ω/K.

Alt text: Scatter plot of the nanocalorimeter's resistance as a function of temperature to calculate the temperature coefficient of resistance.

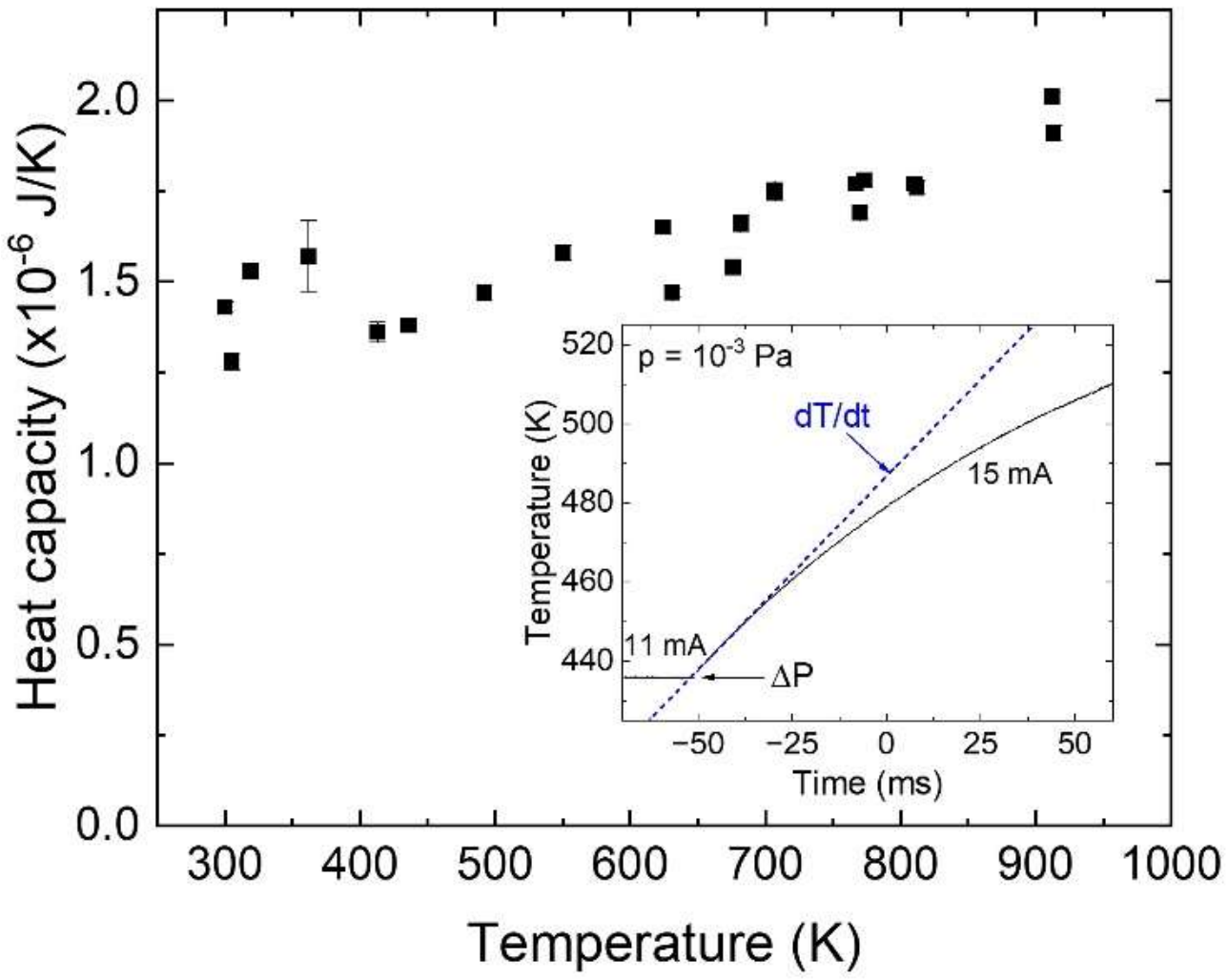

FIG. S2. Evolution of the nanocalorimeter's heat capacity with temperature. Inset: example measurement showing the temperature variation as a current step from 11 mA to 15 mA is applied. The ratio between power variation, $\Delta P$, and time derivative of temperature, $dT/dt$, provides the heat capacity.

Alt text: Scatter plot with the evolution of nanocalorimeter's heat capacity with temperature. An example of heating curve shows the basic parameters to calculate the heat capacity.

## B. Collisional regime of ICP discharge sheaths

To support the analysis of ion energy distribution function (IEDF) measurements, here we discuss the frequency of ion-neutral collisions within the plasma sheath generated by Ar ICP discharges on the grounded retarding field energy analyzer (RFEA). Figure S3 shows pictures of the plasma plumes, striking over the RFEA, captured at 6.5 Pa between 20 W and 80 W RF power. A sub-millimetric sheath contracting as RF power increases can be appreciated by visual inspection.

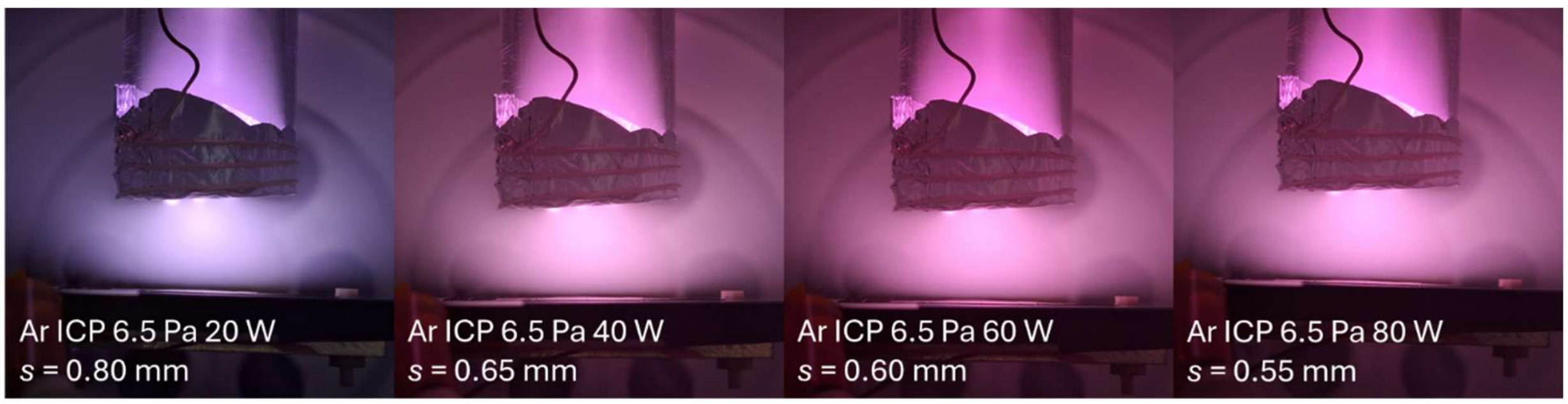

FIG. S3. Images of Ar ICP discharges held at 6.5 Pa between 20 W and 80 W RF over the retarding field energy analyzer (RFEA). The outer diameter of the quartz tube is 25 mm. Sheath thickness, $s$, is calculated using a collisional sheath model [1].

Alt text: Four images of argon glow discharges held between the end of a quartz tube and a planar surface at different RF powers.

The collisional regime is determined by the relation between sheath thickness and ion mean free path. First, sheath thickness, $s$, spanned between 0.55 mm and 0.80 mm as computed from a collisional sheath analysis using the plasma parameters measured via Langmuir probe [1].

Second, mean free path for Ar gas at 6.5 Pa and 300 K is $\lambda_i \approx 1$ mm. Note that this length sets a lower limit for the experimental one: the ion mean free path within the sheath may take higher values due to the reduced cross section acquired by the accelerated ions. Since $\lambda_i > s$ across the explored intervals of pressure and RF power, it can be concluded that the ICP sheaths formed on the RFEA are essentially non-collisional. CCP discharges generate non-collisional anode sheaths as deduced using a similar argument.

### *C. Thermocouple measurements in glow and dark plasma regions*

The images in Figure S4 show the locations where the thermocouple probe measured plasma temperature in argon (Figures S3a,b) and helium (Figures S3c,d) CCP discharges at 53.5 Pa and 60 W RF. The distinct structures observed here owe to the specific plasma dynamics for each element. Ar plasma shows a narrow glow region near the upper electrode (cathode, RF-driven). In contrast, a wider glow region closer to the lower electrode (anode, grounded) is observed for the He discharge.

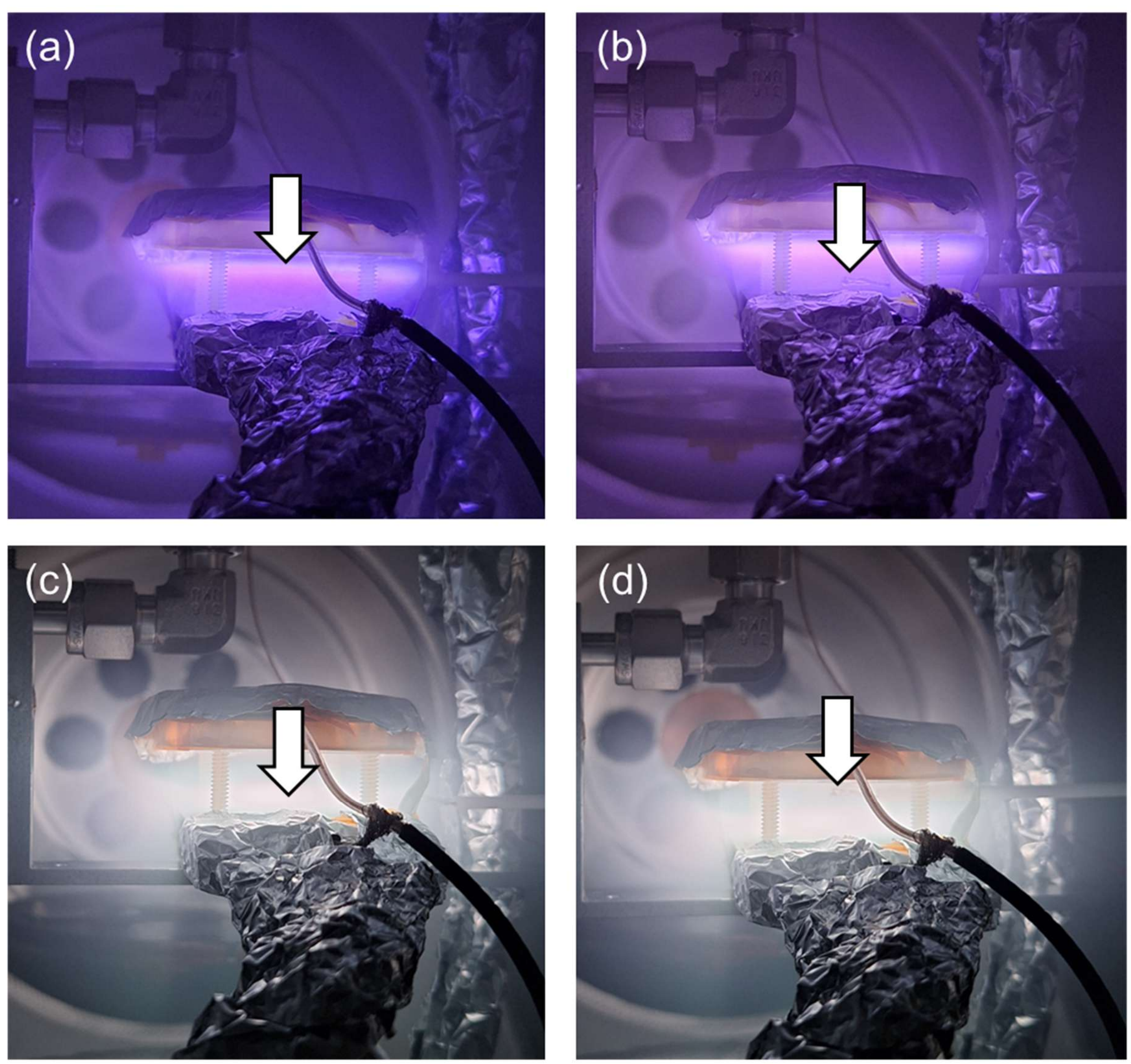

FIG. S4. Images of CCP discharges held at 53.5 Pa and 60 W showing the thermocouple probe located at the axis of the plasma source for Ar in the (a) glow and (b) dark regions, and for He in the (c) glow and (d) dark regions. The arrows indicate the approximate location of the thermocouple. The diameter of the upper, RF-driven electrode is 50 mm.

Alt text: Four images of argon and helium glow discharges and a thermocouple probe located at different axial positions within the discharges.

Temperature variations upon plasma ignition in each case are plotted in Figure S5. In general, monotonous heating tends to saturate with time. As discussed in the main text, temperature increases as the probe radially approaches the center of the CCP cell. Both glow and dark central regions for Ar and He plasmas were explored by shifting the thermocouple vertically along the cell's axis. Higher temperatures were recorded in glow regions in contrast with the dark ones, consistently with stronger gas heating in the bright zones (higher plasma density). The rapid shifts of the signal recorded as soon as plasma was switched ON/OFF may account for electromagnetic disturbances probably due to insufficient shielding of the thermocouple setup.

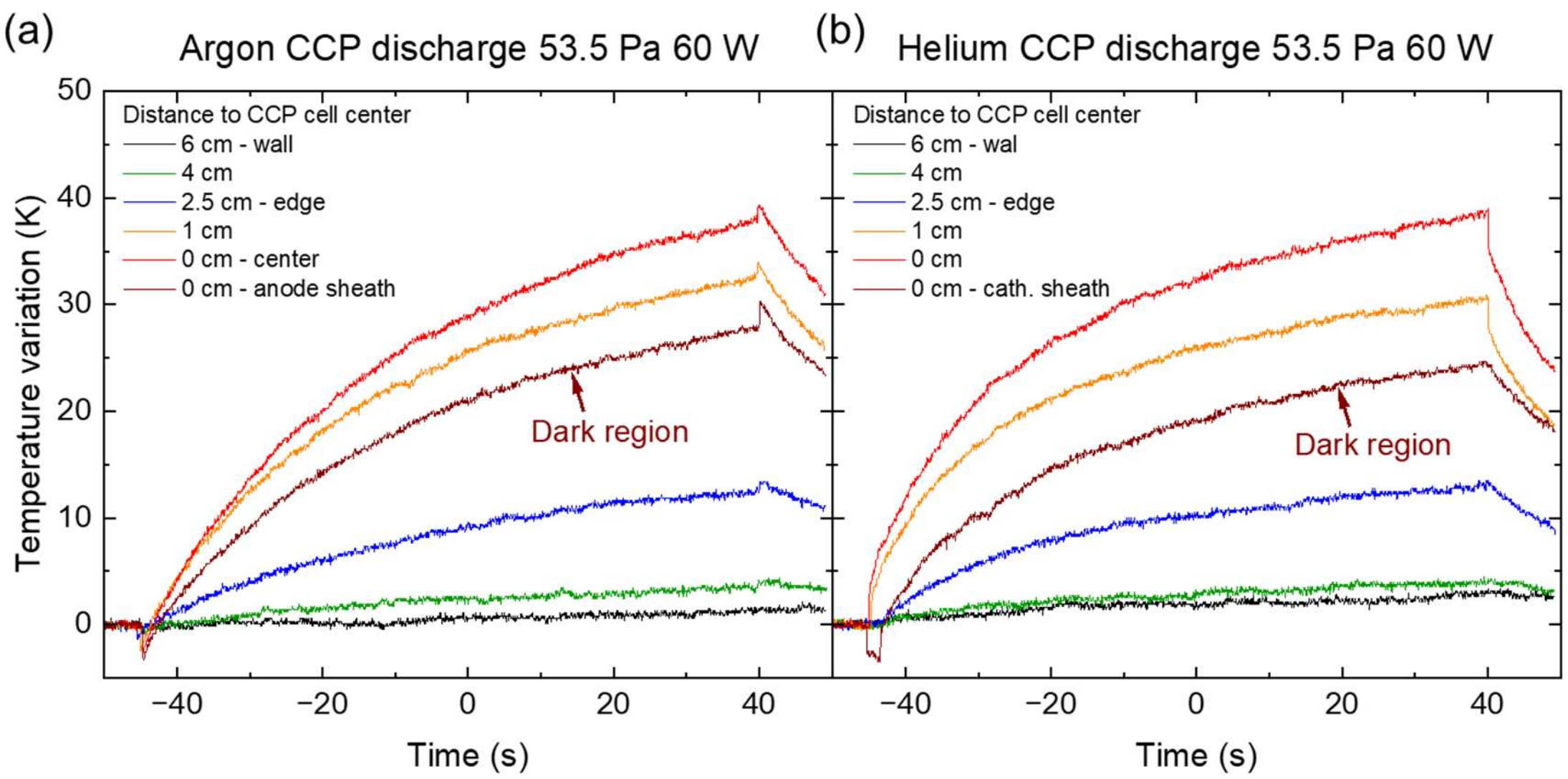

FIG. S5. Temperature response of the thermocouple probe at different locations within the plasma chamber. (a) Ar and (b) He CCP discharges were conducted at 53.5 Pa and 60 W. All measurements correspond to the glow region unless otherwise indicated.

Alt text: Temporal variations of temperature in argon and helium discharges measured with a thermocouple probe located in different radial positions.